\pdfoutput=1
\documentclass[twocolumn,english,aps,pra,showpacs]{revtex4}
\usepackage[T1]{fontenc}
\usepackage[latin1]{inputenc}
\usepackage{color}
\usepackage{amsmath}
\usepackage{amssymb}
\usepackage{graphicx}
\usepackage{esint}
\usepackage{epsfig}
\usepackage{subfigure}

\makeatletter
\@ifundefined{textcolor}{}
{%
 \definecolor{BLACK}{gray}{0}
 \definecolor{WHITE}{gray}{1}
 \definecolor{RED}{rgb}{1,0,0}
 \definecolor{GREEN}{rgb}{0,1,0}
 \definecolor{BLUE}{rgb}{0,0,1}
 \definecolor{CYAN}{cmyk}{1,0,0,0}
 \definecolor{MAGENTA}{cmyk}{0,1,0,0}
 \definecolor{YELLOW}{cmyk}{0,0,1,0}
}

\@ifundefined{definecolor}
 {\usepackage{color}}{}
\@ifundefined{definecolor}{\@ifundefined{definecolor}{\@ifundefined{definecolor}
 {\usepackage{color}}{}
}{}}{}
\usepackage{babel}

\makeatother

\begin{document}

\title{Emergence of Topological and Strongly Correlated Ground States in trapped \\
Rashba Spin-Orbit Coupled Bose Gases}

\author{B. Ramachandhran$^{1}$,  Hui Hu$^{2}$, and Han Pu$^{1}$}

\affiliation{$^{1}$Department of Physics and Astronomy, Rice University, Houston, TX 77005, USA \\
 $^{2}$ARC Centres of Excellence for Quantum-Atom Optics and Centre
for Atom Optics and Ultrafast Spectroscopy, Swinburne University of
Technology, Melbourne 3122, Australia }
\date{\today}

\begin{abstract}
We theoretically study an interacting few-body system of Rashba spin-orbit coupled two-component Bose gases confined in a harmonic trapping potential. We solve the interacting Hamiltonian at large Rashba coupling strengths using Exact Diagonalization scheme, and obtain the ground state phase diagram for a range of interatomic interactions and particle numbers. At small particle numbers, we observe that the bosons condense to an array of topological states with $n+1/2$ quantum angular momentum vortex configurations, where $n = 0, 1, 2, 3...$ At large particle numbers, we observe two distinct regimes: at weaker interaction strengths, we obtain ground states with topological and symmetry properties that are consistent with mean-field theory computations; at stronger interaction strengths, we report the emergence of strongly correlated ground states. 
\end{abstract}
\maketitle
\section{Introduction}
Ultracold atomic gases offer an exceptional platform to explore many-body quantum phenomena due to outstanding experimental control over interatomic interactions, system geometry, density and purity \cite{review_trap_ole}. Numerous research groups have, for example, successfully demonstrated the manifestation of few-body bound states and superfluid states in Bose and Fermi gases in trapped atom experiments  \cite{SF, Effimov}. Furthermore, phenomenal experimental progress has been achieved with atomic gases loaded in optical lattices to emulate traditionally condensed-matter phenomena like superfluid-insulator transition, anti-ferromagnetism, and frustrated many-body systems \cite{MI, AFM, frustrated}. However, due to the neutral nature of atomic gases, most experimental systems were limited to exploring quantum phenomena that would occur in the absence of electromagnetic fields. Recently, even this limitation was overcome, when laser fields were used to successfully generate effective magnetic and electric fields in neutral atoms \cite{IBSemf}. The introduction of (synthetic) gauge fields in ultracold neutral atomic systems has thus opened the possibility of exploring a whole new set of phenomena that would manifest in the presence of abelian and non-abelian vector potentials \cite{gaugefields_review}. 
\par 
In the presence of synthetic gauge fields in trapped ultracold bosonic systems, experimental evidence for spin-orbit (SO) coupling with equal Rashba and Dresselhaus type strengths was reported in a seminal paper \cite{SpielmanNature2011}. Recently, commendable experimental progress has also been achieved towards simulating SO-coupling in ultracold fermionic systems \cite{gauge_fermions}, a phenomenon critical to the simulation of certain topologically insulating states in condensed-matter systems \cite{theory_soc_TI}. In the presence of SO-coupling, a generic Hamiltonian may be broadly classified in two classes: (a) one that breaks $\cal T$ (time-reversal) symmetry, and which can be shown to be gauge-equivalent to a Hamiltonian in the combined presence of abelian and non-abelian vector potentials. For example, authors in Ref.~\cite{IQH} consider an SO-coupling Hamiltonian in the presence of a real (abelian) magnetic field and attempt to simulate the physics of traditional quantum Hall systems; (b) one that preserves $\cal T$ symmetry, and which can be shown to be gauge-equivalent to a Hamiltonian in a pure non-abelian vector potential. In this work, we study an SO-coupling Hamiltonian of the latter class, and discuss the emergence of ground states with unique topological and correlation properties. 
\par
In this manuscript, we study an interacting few-body system of two-component Bose gases confined in a two-dimensional (2D) isotropic harmonic trapping potential with Rashba SO-coupling. The manuscript is organized as follows: In Sec.~\ref{theory}, we outline the model Rashba SO-coupling Hamiltonian and discuss various symmetries. We show that the Hamiltonian is gauge-equivalent to particles subject to a pure non-abelian vector potential that preserves $\cal T$ symmetry. Then, we consider the non-interacting limit of this Hamiltonian, and discuss single-particle solutions at small and large SO-coupling strengths. We proceed to discuss the implementation of Exact Diagonalization (ED) scheme to obtain the low-energy eigenstates of the interacting Hamiltonian in the regime of interest to us - at large SO-coupling strengths. Then, we introduce various analysis techniques,  namely:- energy spectrum, density distribution, single-particle density matrix, pair-correlation function, reduced wavefunction, entanglement spectrum, and entanglement entropy. Each technique would offer its unique perspective to the overall understanding of the ground state properties. 
\par
In Sec.~\ref{secresults}, we discuss the phase diagram and analyze the ground state properties of the interacting Hamiltonian at different particle numbers $N$, and at varied inter-atomic interaction strengths. At small particle numbers with $N=2$, we illustrate the unique topological and symmetry properties of ground states. In the relatively large particle number scenario with $N=8$, we observe that the ground states fall into two distinct regimes: (a) at weak interaction strengths (\emph{mean-field-like  regime}), we observe ground states with topological and symmetry properties that are also obtained via mean-field theory computations; (b) at intermediate to strong interaction strengths (\emph{strongly correlated regime}), we report the emergence of strongly correlated ground states. We proceed to illustrate the topological, symmetry and strong correlation properties of these ground states. Finally in Sec.~\ref{seccon}, we summarize and present concluding remarks.
\section{Theoretical Framework}
\label{theory}
\subsection{System under study}
\label{secham}
We study a two-component Bose gas confined in a 2D isotropic harmonic
trapping potential: $V(\rho)=M\omega_{\perp}^{2}(x^{2}+y^{2})/2=M\omega_{\perp}^{2}\rho^{2}/2$. We consider the Rashba SO-coupling term, that couples pseudo-spin-1/2 degree of freedom and linear momentum, of the form: ${\cal V}_{SO}=-i\lambda_{R}(\hat{\sigma}_{x}\partial_{y}-\hat{\sigma}_{y}\partial_{x})$, where $\lambda_{R}$ is the Rashba SO-coupling strength and $\hat{\sigma}_{x,y,z}$ are $2\times2$ Pauli matrices. The model Hamiltonian for the interacting system is then given by: ${\cal H=}\int d{\bf r}[{\cal H}_{0}+{\cal H}_{{\rm int}}]$, 
\begin{eqnarray}
{\cal H}_{0} & = & \Psi^{\dagger}\left[-\frac{\hbar^{2}\nabla^{2}}{2M}+V\left(\rho\right)+{\cal V}_{SO}-\mu\right]\Psi{\bf ,} \label{H0}\\
{\cal H}_{{\rm int}} & = & (g/2)\sum_{\sigma=\uparrow,\downarrow}\Psi_{\sigma}^{\dagger}\Psi_{\sigma}^{\dagger}\Psi_{\sigma}\Psi_{\sigma}{\bf +}g_{\uparrow\downarrow}\Psi_{\uparrow}^{\dagger}\Psi_{\uparrow}\Psi_{\downarrow}^{\dagger}\Psi_{\downarrow}{\bf ,} \label{Hint}
\end{eqnarray}
where ${\bf r}=(x,y)$ and $\Psi=[\Psi_{\uparrow}({\bf r)},\Psi_{\downarrow}({\bf r)}]^{T}$
denotes the spinor Bose field operators. The chemical potential $\mu$ is to be determined by the total number of bosons $N$ (i.e., $\int d{\bf r}\Psi^{\dagger}\Psi=N$). For simplicity, we have assumed that the intra-component interaction strengths are equal, so that $g_{\uparrow\uparrow}=g_{\downarrow\downarrow}=g$. The Hamiltonian is invariant under symmetry operations associated with the anti-unitary time-reversal operator ${\cal T} = i \hat{\sigma}_{y} {\cal C}$, and the unitary parity operator ${\cal P} = \hat{\sigma}_{z}{\cal I}$, where ${\cal C}$ and ${\cal I}$ perform complex conjugation and spatial inversion operations respectively. The Hamiltonian is also invariant under the combined $\cal PT$ operator, which is unitary since operators ${\cal P}$ and ${\cal T}$ anti-commute, i.e., since $[{\cal P}, {\cal T}]_{+}=0$. We further note that Rashba SO-coupling term breaks inversion symmetry. 
\par
In experiments, the two-dimensionality can be realized by imposing a strong harmonic potential $V(z)=M\omega_{z}^{2}z^{2}/2$ along axial direction in such a way so that $\mu,k_{B}T\ll\hbar\omega_{z}$. For the realistic case of $^{87}$Rb atoms, the interaction strengths can be calculated from the two \textit{s}-wave scattering lengths $a\simeq100a_{B}$ and $a_{\uparrow\downarrow}$, using $g=\sqrt{8\pi}(\hbar^{2}/M)(a/a_{z})$ and $g_{\uparrow\downarrow}=\sqrt{8\pi}(\hbar^{2}/M)(a_{\uparrow\downarrow}/a_{z})$, respectively. Here, $a_{z}=\sqrt{\hbar/(M\omega_{z})}$ is the characteristic oscillator length in $z$-direction, and $a_B$ is the atomic Bohr radius. Note that throughout this work, we consider interaction strengths such that $a_z \gg a, a_{\uparrow\downarrow}$. In another possible regime of strong interactions where $a_z \simeq a, a_{\uparrow \downarrow}$, one needs to include confinement-induced resonance in the calculation of 2D interaction strengths $g$ and $g_{\uparrow\downarrow}$ \cite{PetrovPRL2000}.
\par
In harmonic traps, it is natural to use the trap units; that is, to take
$\hbar\omega_{\perp}$ as the unit for energy, and the harmonic oscillator
length $a_{\perp}=\sqrt{\hbar/(M\omega_{\perp})}$ as the unit for
length. This is equivalent to setting $\hbar=k_{B}=M=\omega_{\perp}=1$.
For the SO-coupling, we introduce an SO-coupling length $a_{\lambda}=\hbar^{2}/(M\lambda_{R})$ and consequently define a dimensionless SO-coupling strength $\lambda_{SO}=a_{\perp}/a_{\lambda}=\sqrt{(M/\hbar^{3})}\lambda_{R}/\sqrt{\omega_{\perp}}$.
In a recent experiment \cite{SpielmanNature2011}, a spinor (spin-1) Bose gas of $^{87}$Rb atoms with $F=1$ ground state electronic manifold is used to create SO-coupling, where two internal "spin" states are selected from this manifold and labelled as pseudo-spin-up and pseudo-spin-down. This gives an effective spin-$1/2$ Bose gas. In this SO-coupled spin-1/2 BEC, $\lambda_{SO}$ is about $10$. In a typical experiment for 2D spin-1/2 $^{87}$Rb BECs \cite{Dalibard2D}, the interatomic interaction strengths are about $g(N-1)\approx g_{\uparrow\downarrow}(N-1)=10^{2}\sim10^{3} (\hbar\omega_{\perp}a_{\perp}^{2})$. These coupling strengths, however, can be precisely tuned by properly
choosing the parameters of the laser fields that lead to the harmonic
confinement and the SO-coupling.
\subsection{Gauge-equivalent form of ${\cal H}_{0}$}
\label{secgauge}
A generic single-particle Hamiltonian may be written in the form ${\cal H}_{g} = ({\bf p} - {\bf A})^2/2M$, where ${\bf p}=\hbar {\bf k}$ is the particle momentum and ${\bf k}$ is the wave-vector. The vector potential ${\bf A}$ may possibly have components in both physical space and spin space. Depending upon the commutation properties of the components of ${\bf A}$, we may hence have an abelian or non-abelian type vector potential. The primary motivation behind deriving a gauge-equivalent form is to map our model Hamiltonian ${\cal H}_{0}$ onto ${\cal H}_{g}$, and hence derive the nature of ${\bf A}$. It is conceivable that depending upon the nature of ${\cal H}_{0}$, ${\bf A}$ could comprise of purely abelian components, or purely non-abelian components, or a combination of both. 

\par
In order to map ${\cal H}_{0}$ onto ${\cal H}_{g}$, it suffices to compare ${\cal H}_{g}$ with the terms $-\hbar^{2}\nabla^{2}/2M - i\lambda_{R}(\hat{\sigma}_{x}\partial_{y}-\hat{\sigma}_{y}\partial_{x})$ in ${\cal H}_{0}$. The latter terms may actually be rewritten as $\left|{\bf p}\right|^{2}/2M + \lambda_{R}(\hat{k}_y \hat{\sigma}_{x} - \hat{k}_x \hat{\sigma}_{y})$. For a two-component Bose gas confined in a 2D isotropic harmonic trap, we have a two-component vector potential ${\bf A}$, with $A_x, A_y$ being $2\times2$ matrices. Comparing ${\cal H}_{0 }$ with ${\cal H}_{g}$, we expect  $A_x \propto \hat{\sigma}_{y}$ and $A_y \propto -\hat{\sigma}_{x}$. Specifically, it can be shown that the vector potential is ${\bf A} = (A_x, A_y, 0) = (\hbar M \omega_{\perp})^{1/2} \lambda_{SO} (\hat{\sigma}_{y}, -\hat{\sigma}_{x}, 0)$. In trap units, we then simply have ${\bf A} = \lambda_{SO} (\hat{\sigma}_{y}, -\hat{\sigma}_{x}, 0)$. The term involving $\left|{\bf A}\right|^{2}$ is a constant, and can be gauged out without loss of generality. Therefore, the strength of the non-abelian vector potential proportionally determines the strength of SO-coupling. It is further evident that $[A_x, A_y] \neq 0$, and that ${\bf A}$ is a pure non-abelian vector potential. Furthermore, the ${\cal T}$ operator commutes with the SO-coupling term $\lambda_{R}(\hat{k}_y \hat{\sigma}_{x} - \hat{k}_x \hat{\sigma}_{y})$. In essence, the model Rashba SO-coupling Hamiltonian in Eqn.~(\ref{H0}) is gauge-equivalent to particles subject to a pure non-abelian vector potential that preserves $\cal T$ symmetry. Proposals to realize vector potentials of similar forms have been addressed by multiple groups \cite{gaugefields_review, gaugeprop1, LLwu, exp_dyn}.
\subsection{Single-particle solutions}
\label{secsps}
We solve the model Hamiltonian $\cal H$ in the absence of interatomic interactions and obtain the single-particle solutions. Rewriting the ${\cal H}_0$ component in Eqn.~(\ref{H0}), the single-particle wavefunction $\phi({\bf r}) = [\phi_{\uparrow}\left({\bf r}\right),\phi_{\downarrow}\left({\bf r}\right)]^{T}$ with energy $\epsilon$ is given by 
\begin{equation}
\left[\begin{array}{cc}
{\cal H}_{osc} & -i\lambda_{R}(\partial_{y}+i\partial_{x})\\
-i\lambda_{R}(\partial_{y}-i\partial_{x}) & {\cal H}_{osc}
\end{array}\right]\left[\begin{array}{c}
\phi_{\uparrow}\\
\phi_{\downarrow}
\end{array}\right]=\epsilon\left[\begin{array}{c}
\phi_{\uparrow}\\
\phi_{\downarrow}
\end{array}\right]\text{,}
\label{eqnsps}
\end{equation}
where ${\cal H}_{{\rm osc}}\equiv-\hbar^{2}\nabla^{2}/(2M)+V\left(\rho\right)$. In polar coordinates ($\rho,\varphi$), we have $-i(\partial_{y}\pm i\partial_{x})=e^{\mp i\varphi}[\pm\partial/\partial\rho-(i/\rho)\partial/\partial\varphi]$. The single-particle wavefunction takes the form 
\begin{equation}
\phi_{m}({\bf r})=\left[\begin{array}{c}
\phi_{\uparrow}(\rho)\\
\phi_{\downarrow}(\rho)e^{i\varphi}
\end{array}\right]\frac{e^{im\varphi}}{\sqrt{2\pi}},
\label{eqnspstates}
\end{equation}
with well-defined total angular momentum $j_{z}$, that is a sum of orbital and spin angular momenta. In general, we may denote the energy spectrum as $\epsilon_{nm}$, where $n=(0,1,2...)$ is the quantum number for the transverse (radial) direction. 
\par
The single-particle wavefunction $\phi_{m}({\bf r})$ is an eigenstate of the unitary ${\cal P}$ operator:
\begin{equation}
{\cal P} \phi_{m}({\bf r}) = \sigma_z (-1)^{m} \left[\begin{array}{c}
\phi_{\uparrow}(\rho)\\
-\phi_{\downarrow}(\rho)e^{i\varphi}
\end{array}\right]\frac{e^{im\varphi}}{\sqrt{2\pi}} = (-1)^{m} \phi_{m}({\bf r}). \nonumber
\end{equation}
The $\cal T$ symmetry preserved by the Hamiltonian results in a two-fold degeneracy (Kramer doublet) of the energy spectrum: any eigenstate $\phi({\bf r})=[\phi_{\uparrow}({\bf r}),\phi_{\downarrow}({\bf r})]^{T}$ is degenerate with its time-reversal partner ${\cal T}\phi({\bf r})= [\phi_{\downarrow}^{*}({\bf r}),-\phi_{\uparrow}^{*}({\bf r})]^{T}$. This symmetry is preserved even in the presence of interatomic interactions, as the terms in interacting Hamiltonian ${\cal H}_{{\rm int}}$ are $\cal T$-invariant. The superposition state, of $\phi_{m}({\bf r})$ and its time-reversal partner state, is an eigenstate of the unitary ${\cal PT}$ operator:
\begin{equation}
{\cal PT} [ \phi_{m}({\bf r}) + {\cal T} \phi_{m}({\bf r}) ] = (-1)^{m+1} [ \phi_{m}({\bf r}) + {\cal T} \phi_{m}({\bf r}) ]. \nonumber
\end{equation}
\begin{figure}[t*]
\begin{centering}
\includegraphics[clip,width=0.48\textwidth]{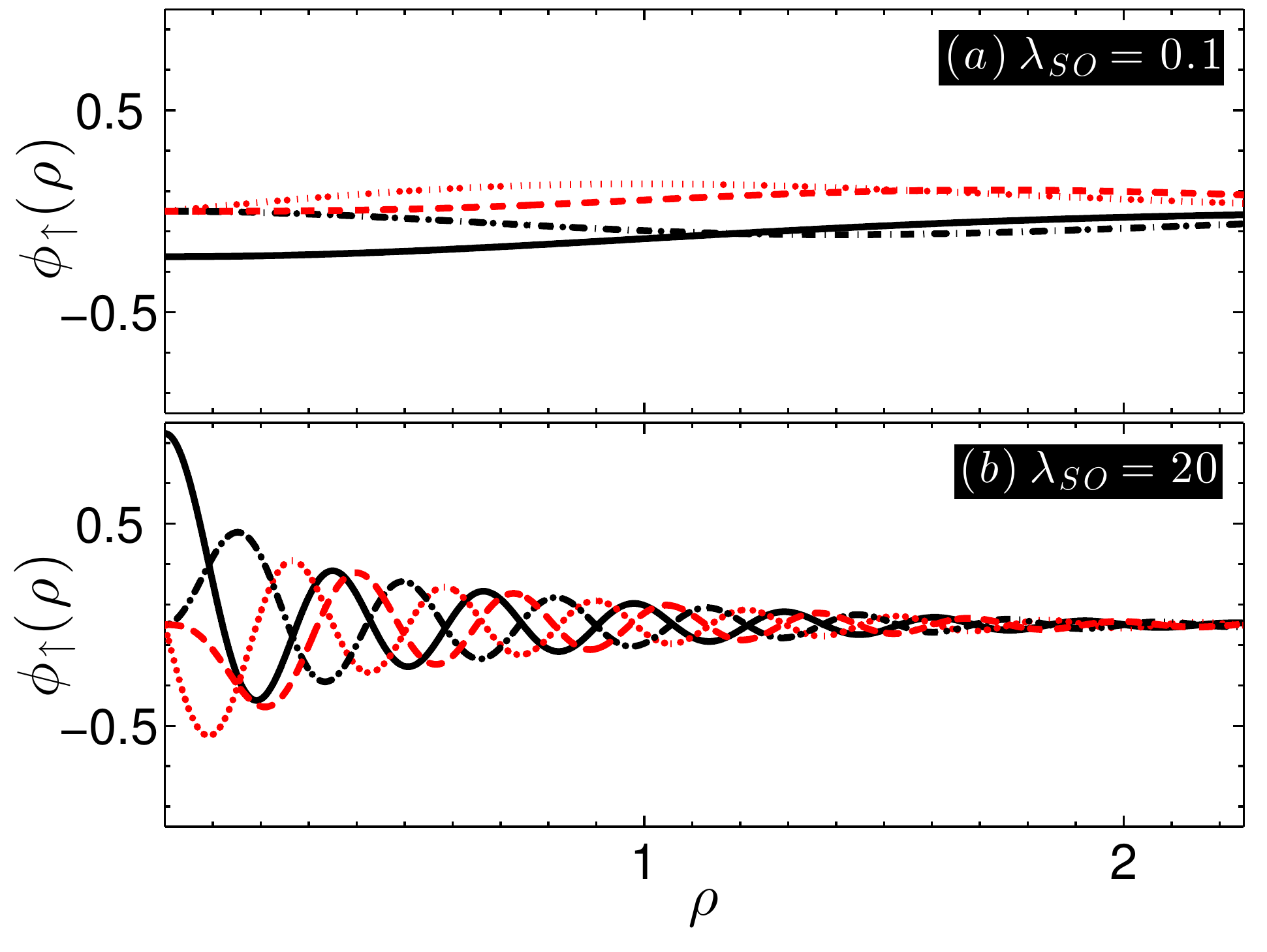}
\par\end{centering}
\caption{(color online). Plots $(a)$ and $(b)$ show wavefunctions $\phi_{\uparrow}(\rho)$ of single-particle states in the $n=0$ manifold at small and large SO-coupling strengths respectively. $m=0$ (solid black), $m=1$ (dotted red), $m=2$ (dash-dotted black) and $m=3$ (dashed red).}
\label{figspsWF} 
\end{figure}
\begin{figure}[t*]
\begin{centering}
\includegraphics[clip,width=0.48\textwidth]{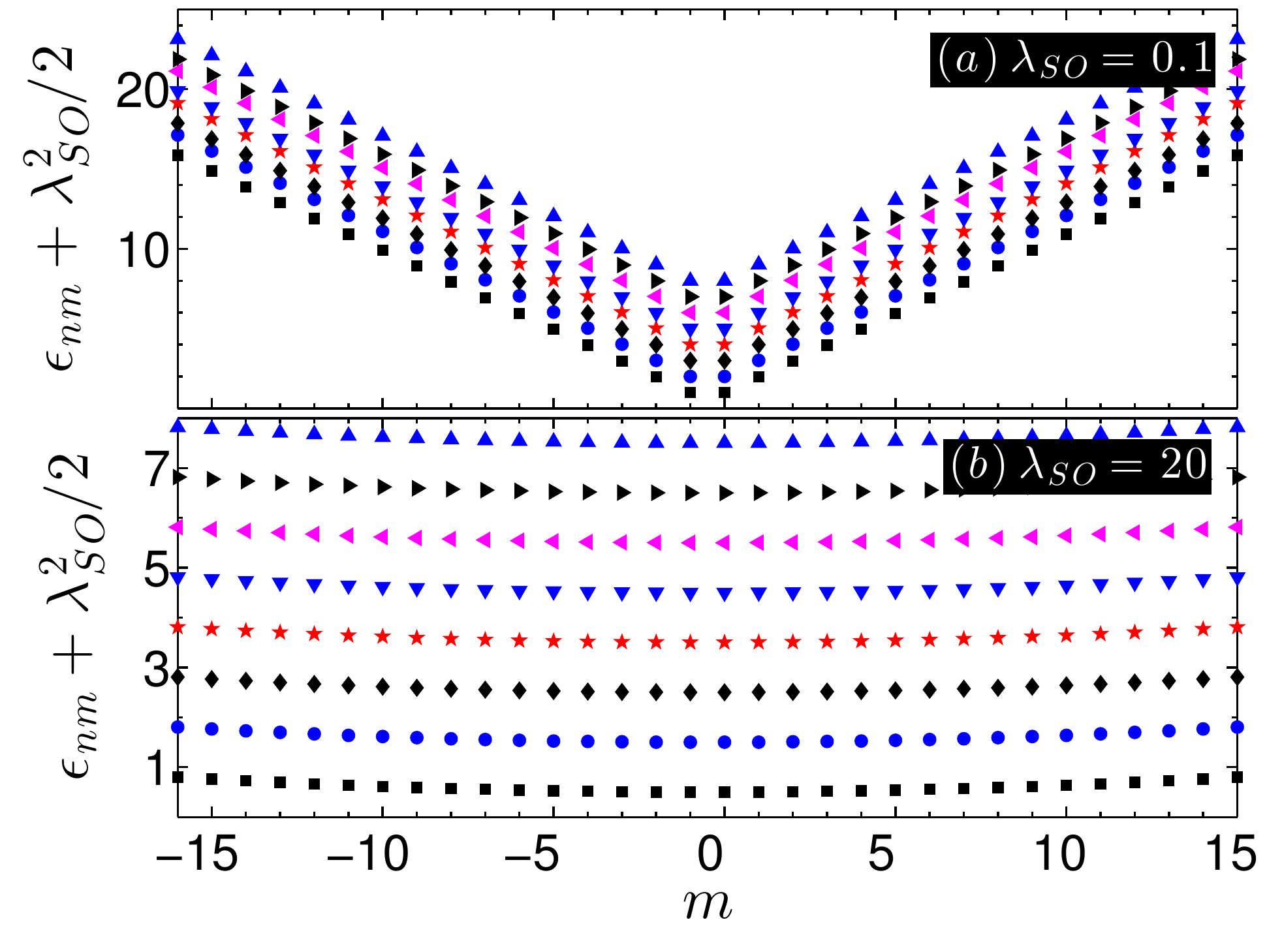} 
\par\end{centering}
\caption{(color online). Plots $(a)$ and $(b)$ show energy spectrum of single-particle states at small and large SO-coupling strengths respectively: $n=0\rightarrow7$ (bottom$\rightarrow$top) and $m=-16\rightarrow+15$. While energies of states within each $n$ are represented by a specific symbol, it is evident that states with higher $n$ have progressively higher energies.}
\label{figspsES} 
\end{figure}
\par
We solve the single-particle spectrum by adopting a numerical basis-expansion method, details of which are outlined in our earlier work \cite{HV12pra}. In Fig.~\ref{figspsWF}, we show wavefunctions of single-particle eigenstates at representative values of small and large SO-coupling strengths. It is evident that a larger SO-coupling strength leads to increased oscillations and increased localization at radii determined by $\left| m \right|$ in the radial direction. Corresponding wavefunctions $\phi_{\downarrow}(\rho)$ also have similar characteristics. In Fig.~\ref{figspsES}, we show the energy spectrum for single-particle states at small and large SO-coupling strengths. From Fig.~\ref{figspsES}$(a)$, it is evident that the energy spectrum is strongly dispersive in $m$ at small SO-coupling strengths, with a large overlap between the energies of single-particle states with different radial quantum number $n$. Qualitatively, the energy spectrum at small SO-coupling strengths may be understood as a weak perturbation of the harmonic oscillator energy levels of the two pseudo-spin components. On the other hand, we observe from Fig.~\ref{figspsES}$(b)$ that the energy spectrum is weakly dispersive or \emph{nearly} flat in $m$ at large SO-coupling strengths. For the range of $m$ shown here, there is no overlap between the energies of single-particle states belonging to different radial quantum numbers $n$, i.e, each $n$ manifold represents single-particle states labelled by their azimuthal angular momenta $m$ with no overlap with adjacent $n$ manifolds. Furthermore, the harmonic trapping potential may be qualitatively understood as a weak perturbation to the energy spectrum at large SO-coupling strengths of the corresponding translationally invariant system. 
\par
The localized nature of the wavefunctions in Fig.~\ref{figspsWF}$(b)$ and the weakly dispersive nature of the single-particle energy spectrum in Fig.~\ref{figspsES}$(b)$ are characteristics that justify a comparison of the single-particle basis states at large SO-coupling strengths with 2D Landau Level (LL) structures in magnetic fields. In Ref.~\cite{LLwu}, the authors discuss the mapping between ${\cal H}_0$ and 2D LL Hamiltonian in a rigorous fashion and generalize the terminology of LLs as `\emph{topological} single-particle level structures labeled by angular momentum quantum numbers with flat or \emph{nearly flat} spectra' \cite{LLwu}. Making use of this generalization, we term the $n=0$ manifold as the lowest LL structure ($LLL$), $n=1$ manifold as the next highest LL, and so on. As seen in Fig.~\ref{figspsES}$(b)$, the radial quantization generates energy gaps between adjacent LLs of the order of trap energy $\hbar \omega_{\perp}$, i.e., of order \emph{unity} in trap units. 
\par
To summarize, we emphasize that the generalized LLs discussed here are created by a truly non-abelian vector potential, i.e., in the absence of any real (abelian) magnetic fields. The strength of Rashba SO-coupling strength, and in-turn the \emph{flatness} of the single-particle energy spectra can be experimentally controlled by using laser fields. At large SO-coupling strengths, as shown for $\lambda_{SO}=20$, we obtain a \emph{nearly flat} single-particle energy spectra. In a non-interacting two-component Bose gas, quantum statistics obviates the occurrence of correlated states in a spectra that is not \emph{perfectly flat}, due to potential condensation of all the particles in the lowest energy single-particle states, identified by $j_z = \pm 0.5$, of the $LLL$ ($n=0$ manifold). However, in the presence of inter-particle interactions, \emph{nearly flat} energy spectra is sufficiently abled to act as an interesting playground to allow for the emergence of strongly correlated ground  states. We now proceed to introduce the ED scheme to solve the interacting Rashba SO-coupled Hamiltonian at large SO-coupling strengths.
\subsection{Interacting few-body problem - Exact Diagonalization scheme}
\label{ED}
We solve the interacting Rashba SO-coupled Hamiltonian $\cal H$ in Eqns.~(\ref{H0}) and (\ref{Hint}) within the Configuration Interaction alias Exact Diagonalization scheme. In this scheme, we expand the interacting many-body Hamiltonian in an appropriate single-particle basis (configuration) to obtain the solution. The solution becomes exact when we consider an infinite number of single-particle states. With $N$ bosons and $M$ single-particle states in the basis, the dimension of Hilbert space is $D = (N+M-1)! / N! (M-1)! $. With $M=24$, for example, $D = 300$ for $N = 2$, and $D = 7888725$ for $N = 8$. The dimension of Hilbert space grows dramatically with system size and hence, for practical purposes, we limit our configuration to a finite size. We observe that the solution becomes \emph{essentially} exact when we consider a \emph{sufficient} number of single-particle states. To solve the problem at hand, it is convenient to work with the SO single-particle basis:
\begin{equation}
\Phi({\bf r})=\sum_{nm} \left[\begin{array}{c}
\phi_{\uparrow nm}({\bf r})\\
\phi_{\downarrow nm}({\bf r})
\end{array}\right] a_{nm} \equiv \sum_{i \equiv nm} \left[\begin{array}{c}
\phi_{\uparrow i}({\bf r})\\
\phi_{\downarrow i}({\bf r})
\end{array}\right] a_{i} \text{,}
\end{equation}
where the field operator $a_i$ is related to the single-particle state $[\phi_{\uparrow nm}({\bf r}), \phi_{\downarrow nm}({\bf r})]^{T}$. Then, Eqns.~(\ref{H0}) and (\ref{Hint}) simply become 
\begin{equation}
{\cal H} = \sum_i \epsilon_i a_i^{\dagger} a_i + \sum_{ijkl} V_{ijkl} a_i^{\dagger} a_j^{\dagger} a_k a_l \text{,}
\label{Hsecquan}
\end{equation}
where $(i,j,k,l)$ collectively denotes $(n,m)$, and $V_{ijkl} = (g/2) [V_{ijkl}^{\uparrow \uparrow} + V_{ijkl}^{\downarrow \downarrow} ]  + g_{\uparrow \downarrow} V_{ijkl}^{\uparrow \downarrow}$ with
\begin{eqnarray}
V_{ijkl}^{\uparrow \uparrow}  & = & \int d\textbf{r} \phi_{\uparrow i}^{*}({\bf r}) \phi_{\uparrow j}^{*}({\bf r}) \phi_{\uparrow k}({\bf r}) \phi_{\uparrow l}({\bf r}) \nonumber \\
V_{ijkl}^{\downarrow \downarrow} & = & \int d\textbf{r} \phi_{\downarrow i}^{*}({\bf r}) \phi_{\downarrow j}^{*}({\bf r}) \phi_{\downarrow k}({\bf r}) \phi_{\downarrow l}({\bf r}) \label{ovlapint}\\
V_{ijkl}^{\uparrow \downarrow}& = & \int d\textbf{r} \phi_{\uparrow i}^{*}({\bf r}) \phi_{\downarrow j}^{*}({\bf r}) \phi_{\uparrow k}({\bf r}) \phi_{\downarrow l}({\bf r}) \text{.}\nonumber
\end{eqnarray}
\par
We perform the ED calculation in Fock space and the Hamiltonian ${\cal H}$ can be written as a matrix of dimension $D^2$, naively accounting for the possibility of inter-coupling every Fock state \cite{EDpaper}. It is clear from the single particle solutions discussed in Eqn.~(\ref{eqnsps}), that the single-particle term $\epsilon_i a_i^{\dagger} a_i$ contributes only to diagonal entries of the Hamiltonian matrix, while the interaction term $V_{ijkl} a_i^{\dagger} a_j^{\dagger} a_k a_l$ contributes to off-diagonal entries as well. The enumeration of off-diagonal entries can be enormously simplified by accounting for a symmetry preserved by ${\cal H}$: conservation of total angular momentum $J_z = \sum_N \, j_z$, as readily seen from Eqn.~(\ref{Hsecquan}). If an entry $V_{ijkl}$ is to be nonzero, we must have $m_i + m_j = m_k + m_l$ in Eqn.~(\ref{ovlapint}). Using only the radial wavefunction, we have (provided $m_i + m_j = m_k + m_l$), 
\begin{eqnarray}
V_{ijkl}^{\uparrow \uparrow}  & = & \frac{1}{2\pi} \int_{0}^{\infty} \rho d\rho \, \phi_{\uparrow i}(\rho) \phi_{\uparrow j}(\rho) \phi_{\uparrow k}(\rho) \phi_{\uparrow l}(\rho) \nonumber \\
V_{ijkl}^{\downarrow \downarrow} & = & \frac{1}{2\pi} \int_{0}^{\infty} \rho d\rho \, \phi_{\downarrow i}(\rho) \phi_{\downarrow j}(\rho) \phi_{\downarrow k}(\rho) \phi_{\downarrow l}(\rho) \\
V_{ijkl}^{\uparrow \downarrow}& = & \frac{1}{2\pi} \int_{0}^{\infty} \rho d\rho \, \phi_{\uparrow i}(\rho) \phi_{\downarrow j}(\rho) \phi_{\uparrow k}(\rho) \phi_{\downarrow l}(\rho) \text{.} \nonumber
\end{eqnarray}
This enables one to visualize the Hamiltonian in \emph{block-diagonal} form, i.e., each \emph{block} is a manifold comprising of Fock states with a fixed $J_z$. Hence, the term $V_{ijkl} a_i^{\dagger} a_j^{\dagger} a_k a_l$ can only couple states within the same manifold, therefore resulting in a sparse Hamiltonian matrix. We solve this sparse matrix to identify the low energy states of the system.
\par
As discussed in Sec.~\ref{secham}, the Hamiltonian $\cal H$ preserves $\cal T$ symmetry. In a certain LL, the energies of states labelled $j_z$ and $-j_z$ are equal and hence, we need to consider both positive and negative angular momentum states in the single-particle configuration. This has two major implications: (a) computational intensity increases tremendously, and (b) a given configuration would never be sufficient to obtain a \emph{complete} $J_z$ manifold, where all contributing single-particle states are included. We note here that the latter issue does not arise when the Hamiltonian breaks $\cal T$ symmetry, as in studies of rotating trapped gases or gases subject to real magnetic fields \cite{Tsymm, Lewenstein06}. In these studies, it was sufficient to consider only positive $j_z$ states and hence obtain \emph{complete} $J_z$ manifolds. In the limit of large SO-coupling strengths, if the interaction strengths are such that the energy contribution from $H_{\textrm{int}}$ is less than unity (in trap units), we may restrict ourselves to the lowest $n=0$ manifold. Within this $LLL$ approximation, we may consider a \emph{sufficient} number of single-particle eigenstates to obtain \emph{essentially exact} low energy eigenstates.
\subsection{Analysis techniques}
\label{secanalysis}
ED scheme enables us to solve the Rashba SO-coupled Hamiltonian $\cal H$ and obtain the ground state phase diagram at various interaction strengths and particle numbers. The ground states have interesting topological, symmetry and strong correlation properties. Here, we outline the details of various techniques that we use to analyze these properties.
\subsubsection{Energy spectrum}
\label{secgses}
First step in our analysis is to identify the total angular momentum manifold $J_z$ to which the ground state belongs. As discussed earlier, the Hamiltonian matrix has a block-diagonal form, with each block identified by its unique $J_z$ value. It is evident that each of these blocks can essentially be diagonalized independently. The energy spectrum comprises of energy eigenvalues from each block, and the lowest eigenvalue and its corresponding $J_z$ may be readily associated with the ground state. Degeneracies in the energy spectrum naturally reflect the degeneracies in the ground state. For example, a typical energy spectrum plot is shown in Fig.~\ref{figN2gses}.
\par
Dimension of Fock space in the ground state $J_z$ manifold will be much smaller when compared to the Hilbert space dimension $D$. For a given parameter set, once we identify the ground state $J_z$ manifold, we can extract the coefficients of all Fock states from the corresponding eigenvector. In essence, we may then represent the ground state wavefunction as a sum of all contributing Fock states: $\Psi_{G} = \sum_{p = 1}^{n_d} \alpha_p \Phi_p$, where $n_d$ is the dimension of ground state $J_z$ manifold and $\alpha_p$ is the coefficient of the Fock state $\Phi_p$. As discussed in Sec.~\ref{secham}, the interacting Hamiltonian ${\cal H}$ is invariant under two \emph{unitary} symmetry operations, ${\cal P}$ and ${\cal PT}$. With the knowledge of ground state wavefunction $\Psi_{G}$, we are now equipped to determine if the ground state is an eigenstate of ${\cal P}$ or ${\cal PT}$ operator.
\subsubsection{Density distribution and single-particle density matrix}
\label{secdd}
With the knowledge of $\Psi_G$, we are equipped to extract various properties of the ground state. We derive density distribution from the expectation value of single-particle density operator, written in second-quantized form as
\begin{equation}
\hat{\rho}(\textbf{r})=\sum_{ij}\langle\phi_i(\textbf{r'})\mid\delta
(\textbf{r} -\textbf{r'})\mid \phi_j(\textbf{r})\rangle a_i^\dag a_j,
\label{eqnddop}
\end{equation}
where $|\phi_i(\textbf{r})\rangle$ is the single-particle state identified by index $j_z$ in the $LLL$ \cite{Lewenstein06}. In our case, we also have an additional index to denote up- and down- spin components. Since $J_z$ is a good quantum number, the operator $a_i^\dag a_j$ selects only one single-particle state within $LLL$ approximation. As a consequence, it does not contain information about products of different amplitudes and loses information about interference pattern \cite{Lewenstein06}. Hence, the density distribution solely preserves the information on individual densities:
\begin{equation}
n(\textbf{r})=\langle \Psi_G\mid\hat{\rho}(\textbf{r})\mid\Psi_G
\rangle= \sum_{i=1}^M\mid\phi_i(\textbf{r})\mid ^2 O_i \:,
\label{eqndd}
\end{equation}
where $O_i$ is the total ground state occupation of the single-particle state
$|\phi_i(\textbf{r})\rangle$ \cite{Lewenstein06}. Within the $LLL$ approximation, $O_i$ are essentially eigenvalues of the diagonal single-particle density matrix. Since single-particle states in Eqn.~(\ref{eqnspstates}) are eigenstates of ${\cal P}$ operator, it is evident that the density distributions $n(\textbf{r})$ would be cylindrically symmetric. For example, representative plots of $O_i$ as a function of $j_z$, and plots of density distributions are shown in Figs.~\ref{figN2all} and \ref{figN8all}.
\subsubsection{Pair-correlation function}
\label{secpcf}
Pair-correlation functions help us analyze the internal structure of the ground states. We write the pair-correlation operator (not normalized) in second-quantized form \cite{Lewenstein06},
\begin{equation}
\hat{\rho}(\textbf{r},\textbf{r}_0)=\sum_{ijkl}
\phi^*_i(\textbf{r}) \phi^*_j(\textbf{r}_0)\phi_k(\textbf{r}) \phi_l(\textbf{r}_0)
a_i^\dag a_j^\dag a_la_k.
\label{eqnpcfop}
\end{equation}
In our case, we also have an additional index to denote up- and down-spin components. For instance, we may compute pair-correlation functions that determine the conditional probability to find an up-spin or a down-spin, when an up-spin component is assumed to be present at a fixed point $\textbf{r}_0$, i.e., $\langle n_{\uparrow}(\textbf{r}_0) n_{\uparrow}(\textbf{r})\rangle$ or $\langle n_{\uparrow}(\textbf{r}_0) n_{\downarrow}(\textbf{r})\rangle$ respectively. We may choose $\textbf{r}_0$ to be away from the origin, but with a substantial amplitude of $n(\textbf{r})$. Due to angular momentum conservation, the condition $i+j=k+l$ must further be fulfilled. Computing the expectation value of $\hat{\rho}(\textbf{r},\textbf{r}_0)$ with respect to $\Psi_G$, we obtain the pair-correlation function as
\begin{eqnarray}
\rho(\textbf{r},\textbf{r}_0) = \sum_{ijkl}\sum_{pp'}&\,&\alpha_p^* \alpha_{p'}
\phi^*_i(\textbf{r}) \phi^*_j(\textbf{r}_0)\phi_k(\textbf{r}) \phi_l(\textbf{r}_0) \nonumber \\
&\,&\langle\Phi_p\mid a_i^\dag a_j^\dag a_la_k\mid \Phi_{p'}\rangle \text{.}
\label{eqnpcf}
\end{eqnarray}
When the wavefunction $\Psi_{G}$ is an eigenstate of ${\cal PT}$ operator, pair-correlation function illustrate the ground state symmetry properties. Furthermore, they reveal the correlations between up- and down-spin components in real-space. Pair-correlation functions at representative interaction strengths are shown in Figs.~\ref{figN2all} and \ref{figN8all}.
\subsubsection{Reduced wavefunction}
\label{seccwf}
We shall now discuss techniques to analyze if the ground states possess vortex structures with distinct topological properties. One identifying property is the presence of quantized values of \emph{skyrmion} number, as discussed in our earlier work \cite{HV12pra}. However, this requires the computation of ground state wavefunction in real-space, a computationally prohibitive task for the bosonic few-particle system under study. Here, we discuss a viable approach to identify the topological nature of the ground state by computing the reduced wavefunction \cite{CWF}: 
\begin{equation}
\psi_{\textrm{rwf}}({\bf r}) = \frac{\Psi({\bf{r},\bf{r_2^*},...,\bf{r_N^*}})}{\Psi({\bf{r_1^*},\bf{r_2^*},...,\bf{r_N^*}})} \text{.}
\label{eqncwfop}
\end{equation}
Reduced wavefunction $\psi_{\textrm{rwf}}({\bf r})$ is computed with respect to one particle, here particle with index 1, while the remaining $N-1$ particles are placed at their most probable locations $\bf{r_i^*}$ \cite{CWF}. In our case, we also have an additional index to denote up- and down-spin components. With $\psi_{\textrm{rwf},\uparrow}({\bf r})$ and $\psi_{\textrm{rwf},\downarrow}({\bf r})$ known, we can now extract phase information and compute a distinct topological quantity, \emph{vorticity}, i.e., the number of phase slips from $+\pi$ to $-\pi$ along a closed contour. An integer-valued vorticity is an unambiguous way of establishing that the ground state is topological in nature with a distinct vortex structure. For example, typical phase plots  revealing different vorticities are shown in Figs.~\ref{figN2all} and \ref{figN8all}. 
\subsubsection{Entanglement measures}
\label{secem}
We compute entanglement measures to analyze correlation properties of various ground states. In particular, we intend to probe the ground state correlation properties that specifically stem from the presence of inter-particle interactions. To achieve this goal, we take cues from seminal papers in Ref.~\cite{ES}. We choose a \emph{proper} single-particle basis comprising of the set of eigenstates in Eqn.~(\ref{eqnspstates}) of the single-particle Hamiltonian ${\cal H}_0$. In such a single-particle basis, entanglement in the ground state, or any non-degenerate energy eigenstate, occurs specifically due to the presence of interactions \cite{ES}. 
\par
The first step in discussing any entanglement measure is to partition the system and compute entanglement properties between different subsystems. As discussed in Sec.~\ref{secsps}, similar to 2D LL orbitals, the single-particle eigenstates at large SO-coupling strengths are fairly localized in nature. This warrants us to consider partitioning the system in \emph{orbital space} \cite{OES}. The $\cal T$ symmetry preserved by the Hamiltonian $\cal H$ naturally prompts us to partition the orbitals into two subsystems: positive $j_z$ states (subsystem $A$) and negative $j_z$ states (subsystem $B$). We write the ground state wavefunction in Fock space as $\Psi_{G} = \sum_{p = 1}^{n_d} \alpha_p \Phi_p$, where $\Phi_p$ is represented as $\mid n_{-j_c} n_{-(j_c-1)} .. .. n_{j_c-1} n_{j_c} \rangle$. Here, $n_{j_z}$ represents the occupation number of the single-particle eigenstate $j_z$, and as discussed in Sec.~\ref{ED}, a finite size cut-off is made at a certain value $j_{c} \equiv j_{z,c}$ for computational feasibility. Now, we proceed to compute the bipartite entanglement properties between subsystems $A$ and $B$, i.e., between the positive and negative $j_z$ states respectively.
\par
\emph{Orbital entanglement spectrum:-} With the knowledge of $\Psi_{G}$, we compute the entries of the density matrix $\hat{\rho}$ for the ground state as 
\begin{equation}
\langle n_{-j_c}^{'} .. .. n_{j_c}^{'} \mid \hat{\rho} \mid n_{-j_c} .. .. n_{j_c} \rangle = \alpha_p \alpha_p^* \text{,}
\label{eqndmop}
\end{equation}
\noindent where the generic density operator is $\hat{\rho} = \mid \Psi_{G} \rangle \langle \Psi_{G} \mid$. 
\par
Now, we compute the reduced density matrix (RDM) $\hat{\rho}_A$ by tracing out the degrees of freedom of subsystem $B$, meaning $\hat{\rho}_A$=Tr$_B \, \hat{\rho}$. As shown in Ref.~\cite{ES}, occupation numbers act as distinguishable degrees of freedom in characterizing entanglement in a finite system of identical quantum particles. Hence in our study, RDM is computed by tracing out the occupation of all the negative $j_z$ states from the density matrix:
\begin{eqnarray}
&\,& \langle n_{1/2}^{'} .. .. n_{j_c}^{'} \mid \hat{\rho}_{j_c}(1/2,..,j_c) \mid n_{1/2} .. .. n_{j_c} \rangle = \\
&\,& \sum_{n_{-j_c} ..n_{-1/2}} \langle n_{-j_c} ..n_{-1/2} n_{1/2}^{'}.. n_{j_c}^{'} \mid \hat{\rho} \mid n_{-j_c} ..n_{-1/2} n_{1/2}.. n_{j_c}  \rangle \nonumber 
\label{eqnrdmop}
\end{eqnarray}
The RDM $\hat{\rho}_A$ has a block-diagonal structure, with each block characterized by the total angular momentum $J_z^A$ that corresponds only to particles in subsystem $A$. The block-diagonal structure allows us to compute all the eigenvalues of the RDM using full-diagonalization techniques. Orbital entanglement spectrum ($OES$), termed so because the partition is defined in orbital space, is the plot of entanglement pseudo-energies $\xi_i$ as a function of $J_z^A$. Here, 
$\xi_i = - \textrm{ln} \, \rho_i^A$, with $\rho_i^A$ being the eigenvalues of RDM $\hat{\rho}_A$ \cite{Haldane}. It is evident that $\xi_i$ with smaller magnitudes maximally contribute to the ground state properties. 
\par
Plots of $OES$ reveal information about the occupation of various Fock states in a given ground state manifold, and in-turn the correlation properties of the ground state. If various Fock states $\Phi_p$ in the ground state $J_z$ manifold have similar magnitudes of $\alpha_p$, it results in similar RDM eigenvalues of  $\rho_i^A$, and in-turn, similar magnitudes of $\xi_i$. Thus, if an $OES$ plot reveals that $\xi_i$ values are degenerate or \emph{nearly} degenerate, this is a clear manifestation of the correlated nature of the ground state. On the other hand, if the $OES$ plot reveals that the values of $\xi_i$ are distinctly non-degenerate, the ground state is clearly \emph{not} correlated. For example, representative $OES$ plots are shown in Figs.~\ref{figN2all}, and \ref{figN8all}.
\par
\emph{Entanglement entropy:-} Plots of $OES$ reveal the whole spectrum of eigenvalues of the RDM and help us understand the correlation properties of the ground state. However, it is sometimes useful to extract just a single representative quantity from the RDM \cite{HaqueEE}. Entanglement entropy ($EE$) is such a measure that can be readily obtained from the set of eigenvalues $\rho_i^A$ of the RDM $\hat{\rho}_A$, and is defined as $S_A=-\textrm{tr}[\hat{\rho}_A \, \textrm{ln} \hat{\rho}_A]=-\sum_i \rho_i^A \, \textrm{ln} \rho_i^A$. A higher entropy value means that the ground state is more homogeneously spread in Fock space, i.e., a larger number of Fock states $\Phi_p$ make substantial contributions towards the ground state. A distinct advantage of an $EE$ plot is that we are able to look at entropy values for a whole range of interaction strengths in a single plot, and thereby, understand correlation properties of various phases. For example, representative $EE$ plots are shown in Figs.~\ref{figN2ee05}, \ref{figN2ee15}, \ref{figN8ee05}, and \ref{figN8ee15}.
\par
In summary, density distribution, eigenvalues of single-particle density matrix, pair-correlation function and reduced wavefunction would help us identify various symmetry and topological properties of the ground states. Computation of RDM from \emph{proper} single-particle basis enables us to extract various entanglement measures and allow us to analyze correlation properties that specifically stem from inter-particle interactions.

\section{Results and Discussion}
\label{secresults}
As discussed in Sec.~\ref{secsps}, in the absence of interactions, all particles would simply condense into the two lowest energy single-particle eigenstates in the $LLL$ identified by quantum numbers $j_z = \pm 0.5$. This is due to the weak, but finite, dispersion in $j_z$ present in the single-particle energy spectrum shown in Fig.~\ref{figspsES}$(b)$. The ${\cal P}$-eigenstate, identified by $j_z=+0.5$, is represented by wavefunction ${\Phi}_{\cal P} = [\phi_{\uparrow}(\rho),\phi_{\downarrow}(\rho)e^{i\varphi}]^{T}/\sqrt{2\pi}$. It has a half-quantum vortex configuration, as the spin-up component stays in the $s$-state and the spin-down component is in the $p$-state \cite{HV12pra, WuCPL2011, HQVS}. The resulting spin texture of this topological state is of skyrmion type \cite{HV12pra}. The degenerate time-reversed ${\cal P}$-eigenstate, identified by $j_z=-0.5$ and represented by ${\cal T}{\Phi}_{\cal P}=[\phi_{\downarrow}(\rho)e^{-i\varphi},-\phi_{\uparrow}(\rho)]^{T}/\sqrt{2\pi}$, also has a half-quantum vortex configuration. We may as well construct a zero angular momentum ${\cal PT}$-eigenstate, from an equal superposition of opposite angular momentum ${\cal P}$-eigenstates: $\Phi_{{\cal PT},j_z=0}  = \left( {\Phi}_{\cal P} \pm {\cal T} \, {\Phi}_{\cal P}\right)/\sqrt{2}$. In the absence of interactions, either of the ${\cal P}$-eigenstates or the superposition ${\cal PT}$-eigenstate are degenerate. In addition, any arbitrary superposition of the degenerate ${\cal P}$-eigenstates, which in principle need not be a ${\cal PT}$-eigenstate, will also be a degenerate ground state. 
\par
In the presence of inter-particle interactions, the ground state is not anymore determined solely by the energy contribution of the non-interacting part of the Hamiltonian ${\cal H}_{0}$.  Depending upon the strengths of $g$ and $g_{\uparrow\downarrow}$, the energy contribution from the interacting part of the Hamiltonian ${\cal H}_{{\rm int}}$  also plays a crucial role. This competition can be better understood, especially at large SO-coupling strengths, by analyzing the single-particle wavefunctions and energy. As shown in Fig.~\ref{figspsES}$(b)$, energy contributions due to ${\cal H}_{0}$ tries to keep the particles in states with lower value of angular momenta $j_z$. However, for repulsive interaction strengths, energy considerations due to ${\cal H}_{{\rm int}}$ tries to keep the particles as far away from each other as possible. This in-turn means that the particles tend to occupy states with larger value of angular momenta, since they have a larger localization radii as shown in Fig.~\ref{figspsWF}$(b)$. In essence, the ground state of the interacting many-body Hamiltonian is determined by the competition between the ${\cal H}_{0}$ and ${\cal H}_{{\rm int}}$ terms. 
\par
The simplest scenario where the competition between the ${\cal H}_{0}$ and ${\cal H}_{{\rm int}}$ terms, in-turn the effect of inter-particle interactions, clearly manifests is in an interacting problem with $N=2$ particles. For this reason, we discuss the results for $N=2$ particles and analyze the ground state properties in greater detail, before proceeding to larger particle numbers. We solve the interacting few-body Hamiltonian $\cal H$ at large SO-coupling strengths using ED scheme within $LLL$ approximation. The computational intensity, especially at large interaction strengths, limits the feasibility of this scheme to the order of $N=8$ particles \cite{quadrops}. In an earlier mean-field study on \emph{homogeneous} two-component Bose gas \cite{ZhaiPRL2010}, it was shown that the particles condense into either a single plane-wave state (for $g>g_{\uparrow\downarrow}$) or a density-stripe state (for $g<g_{\uparrow\downarrow}$). Similarly, in our earlier related work on \emph{trapped} two-component Bose gas \cite{ourPRL}, depending on the relative magnitudes of $g$ and $g_{\uparrow\downarrow}$, we show that states with distinct topological and symmetry properties emerge in the mean-field phase diagram. Taking cues from these results, in this study, we solve for the ground state wavefunction at various interaction strengths, however fixing the relative magnitude $g_{\uparrow\downarrow}/g$ at 0.5 or 1.5. In this section, we present the results at different particle numbers $N$, and analyze the topological, symmetry and correlation properties of the ground states using various techniques discussed in Sec.~\ref{secanalysis}. 
\subsection{$N=2$}
\label{secn2}
\begin{figure}[t*]
\centering
\begin{tabular}{cc}
\includegraphics[clip,width=0.48\textwidth]{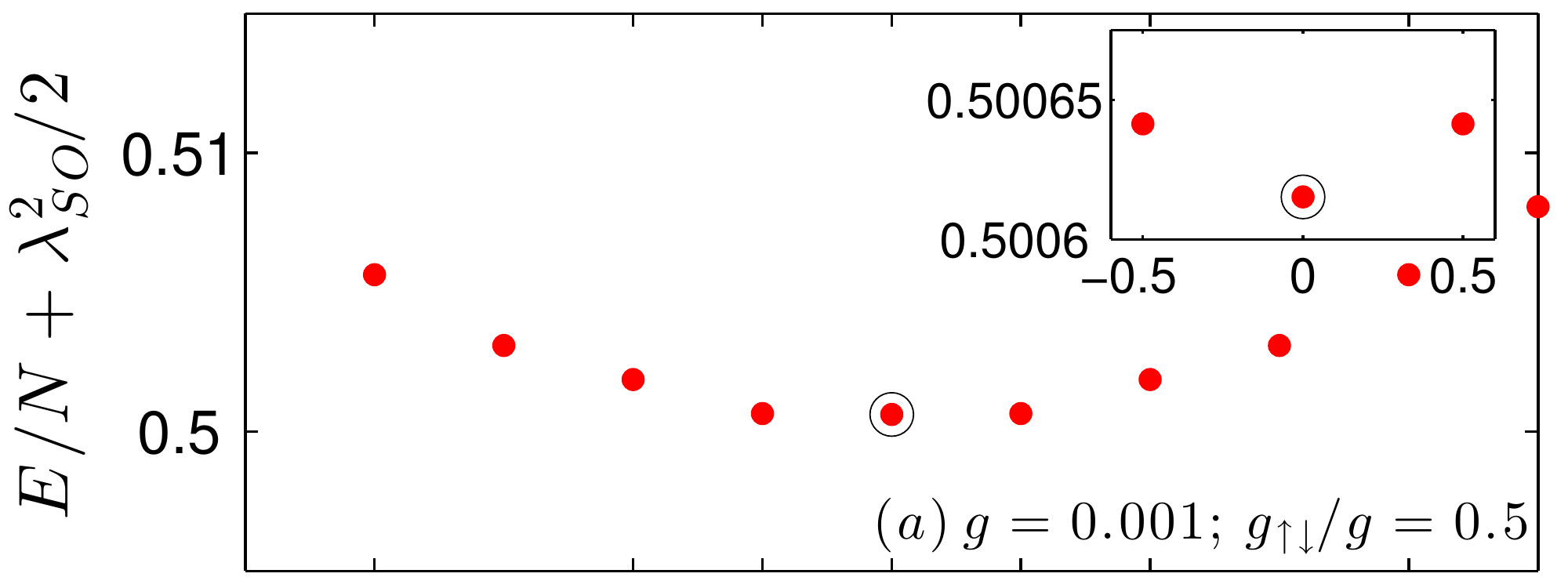} \\
\includegraphics[clip,width=0.48\textwidth]{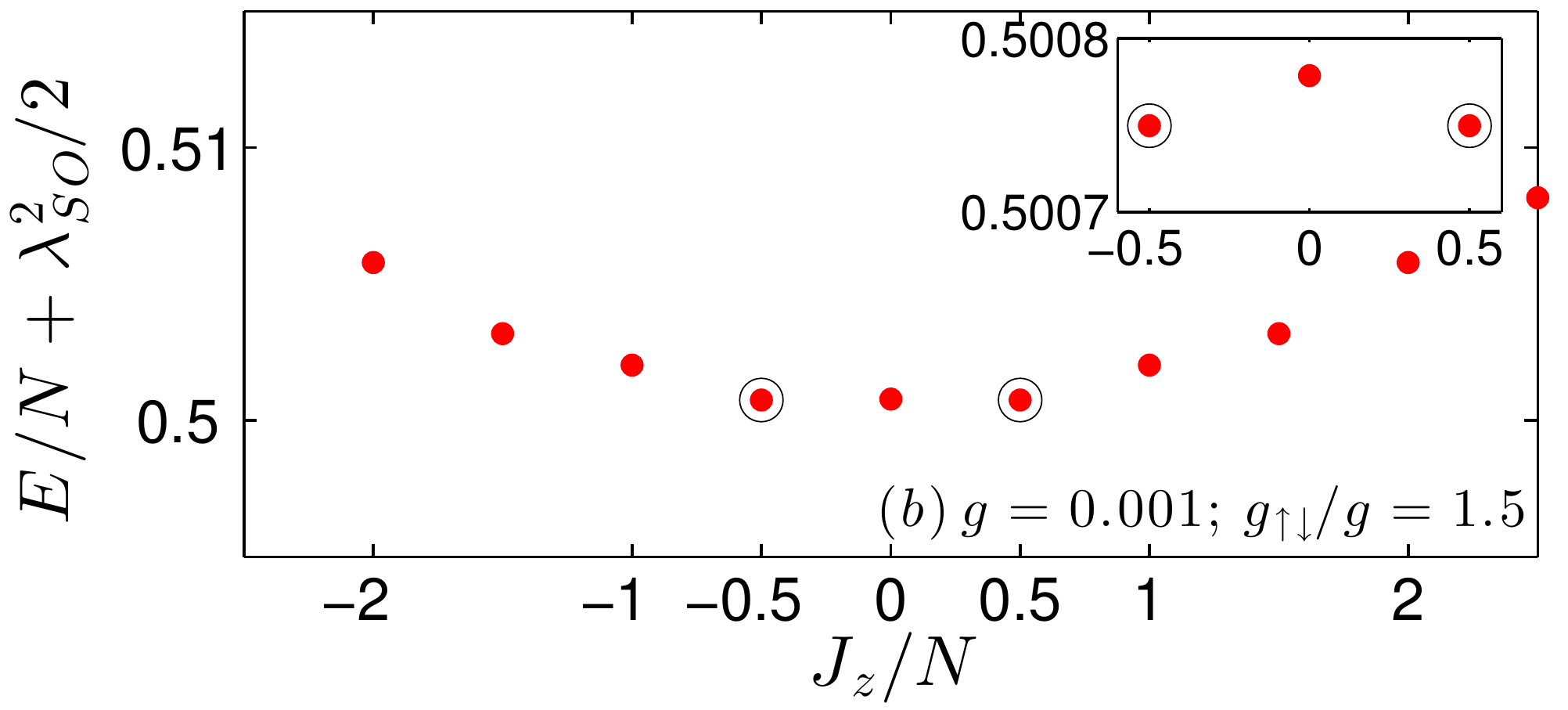} 
\end{tabular}
\caption{(color online). Energy spectrum for extremely weak interaction strengths with $\lambda_{SO}=20$ and $N = 2$. Here, each marker (red) represents the lowest energy eigenvalue of a specific \emph{block diagonal} with a fixed value of $J_z$. Since energy eigenvalues are very close, we identify the ground state energies by circled (black) markers and further, show the zoomed-in plots in the inset.}
\label{figN2gses}
\end{figure}
\begin{figure}[t!]
\begin{centering}
\includegraphics[clip,width=0.48\textwidth]{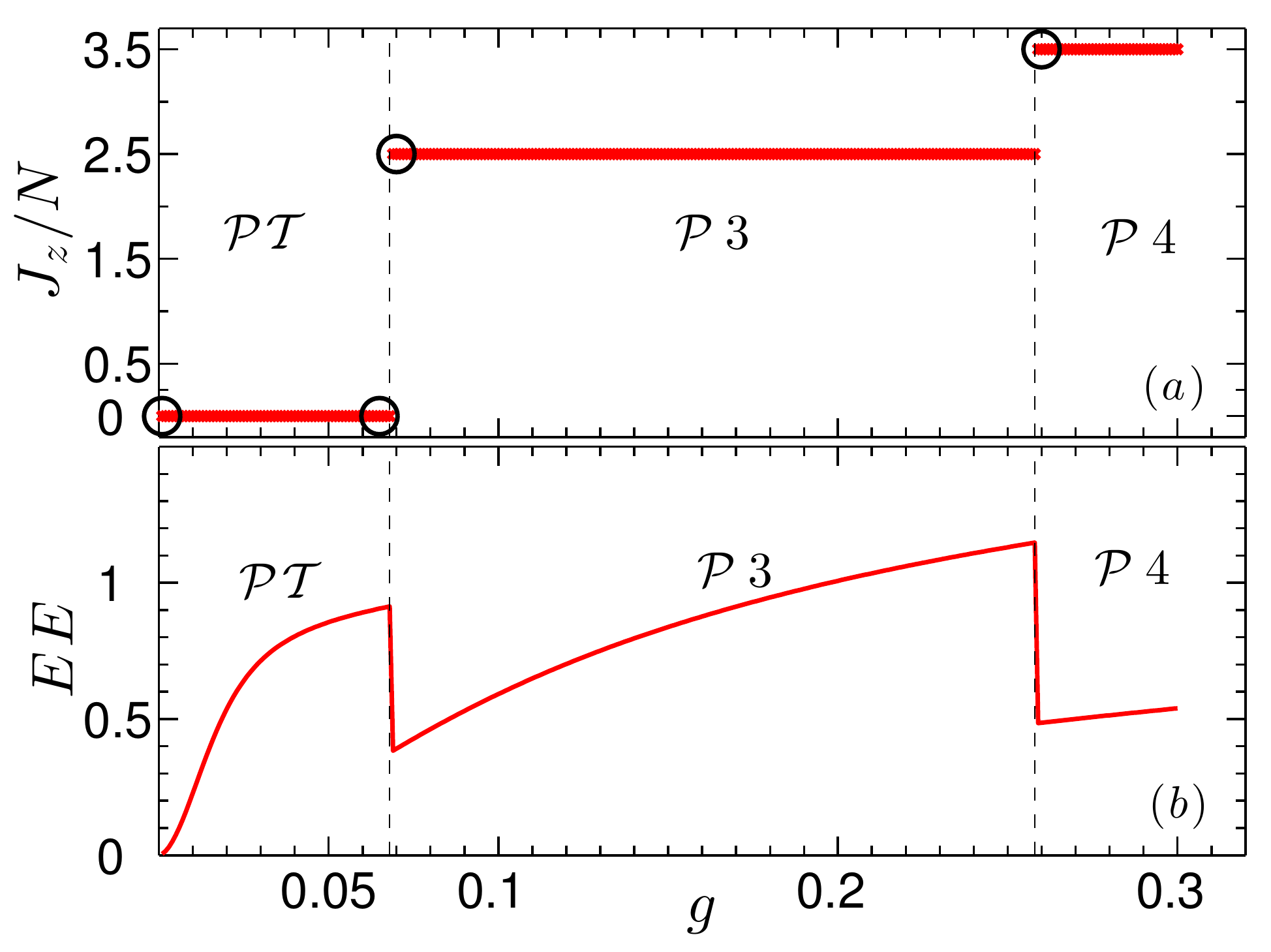} 
\par\end{centering}
\caption{(color online). Plots of $(a)$ ground state $J_z/N$ manifolds and $(b)$ entanglement entropy, as a function of interaction strength $g$ with $\lambda_{SO}=20, N = 2, g_{\uparrow\downarrow}/g=0.5$. For representative interaction strengths denoted by circled (black) markers, we illustrate the ground state properties in Fig.~\ref{figN2all}.}
\label{figN2ee05} 
\end{figure}
\par
As discussed in Sec.~\ref{secgses}, we analyze the energy spectrum to identify the ground state angular momentum manifold $J_z$, or equivalently, $J_z/N$. In Fig.~\ref{figN2gses}$(a)$, we notice that the ground state belongs to $J_z/N=0$ manifold. We further determine that the ground state wavefunction $\Psi_G$ is an eigenstate of ${\cal PT}$ operator. On the other hand, we observe from Fig.~\ref{figN2gses}$(b)$ that the ground state is degenerate in $J_z/N= \pm0.5$ manifolds. In either scenario, in Fig.~\ref{figN2gses}$(b)$, we determine that $\Psi_G$ is an eigenstate of ${\cal P}$ operator. It is evident that, even in the presence of extremely weak interaction strengths, the interacting Hamiltonian picks either a ${\cal P}$-eigenstate or a ${\cal PT}$-eigenstate to be the ground state. Furthermore, it is clear that the ground state is sensitive to the relative magnitudes of $g_{\uparrow\downarrow}$ and $g$. 
\begin{figure}[t!]
\begin{centering}
\includegraphics[clip,width=0.48\textwidth]{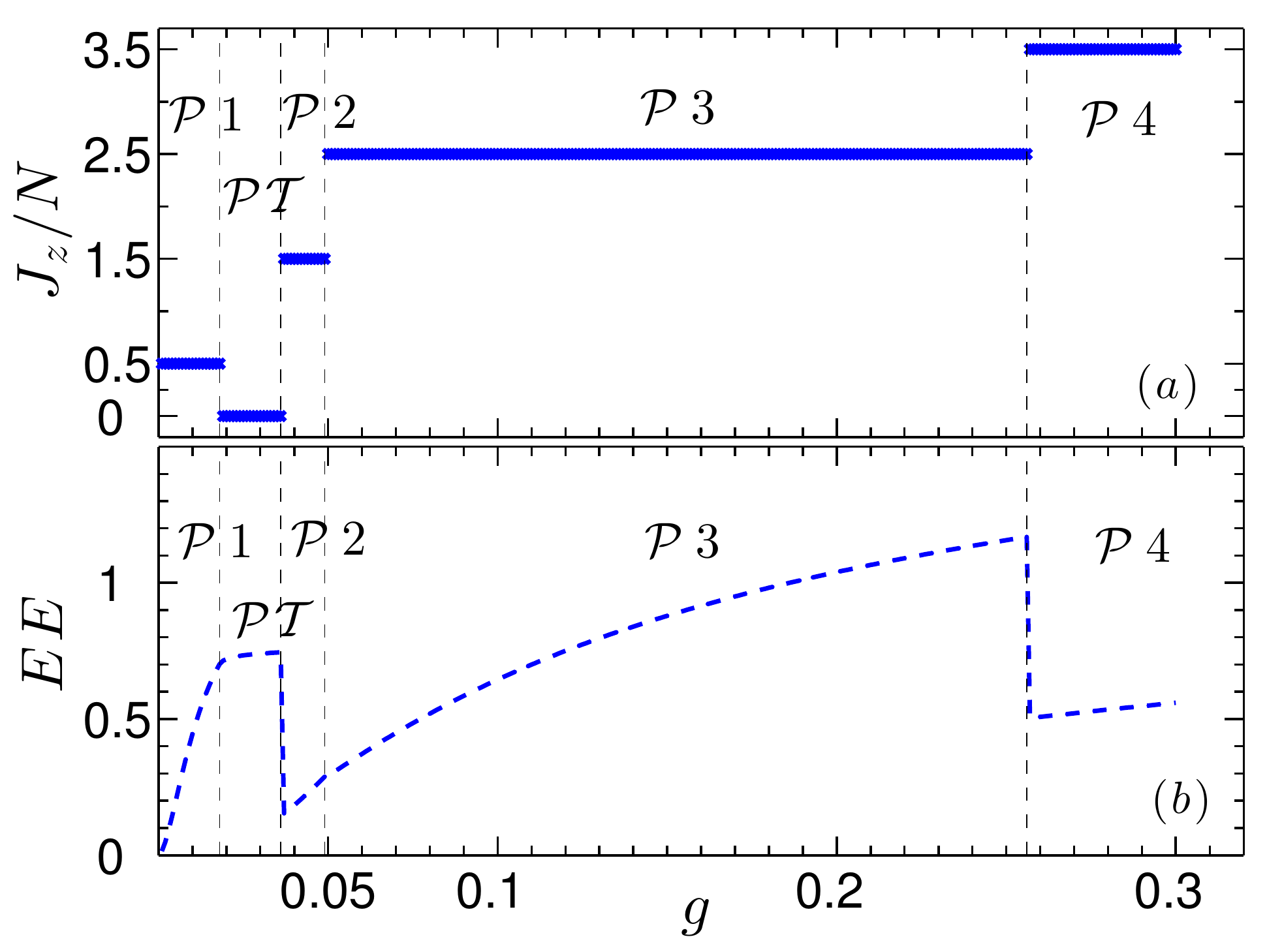} 
\par\end{centering}
\caption{(color online). Plots of $(a)$ ground state $J_z/N$ manifolds and $(b)$ entanglement entropy, as a function of interaction strength $g$ with $\lambda_{SO}=20, N = 2, g_{\uparrow\downarrow}/g=1.5$.}
\label{figN2ee15} 
\end{figure}
\begin{figure*}[ht!]
\centering
\begin{tabular}{cccc}
\includegraphics[clip,width=0.225\textwidth]{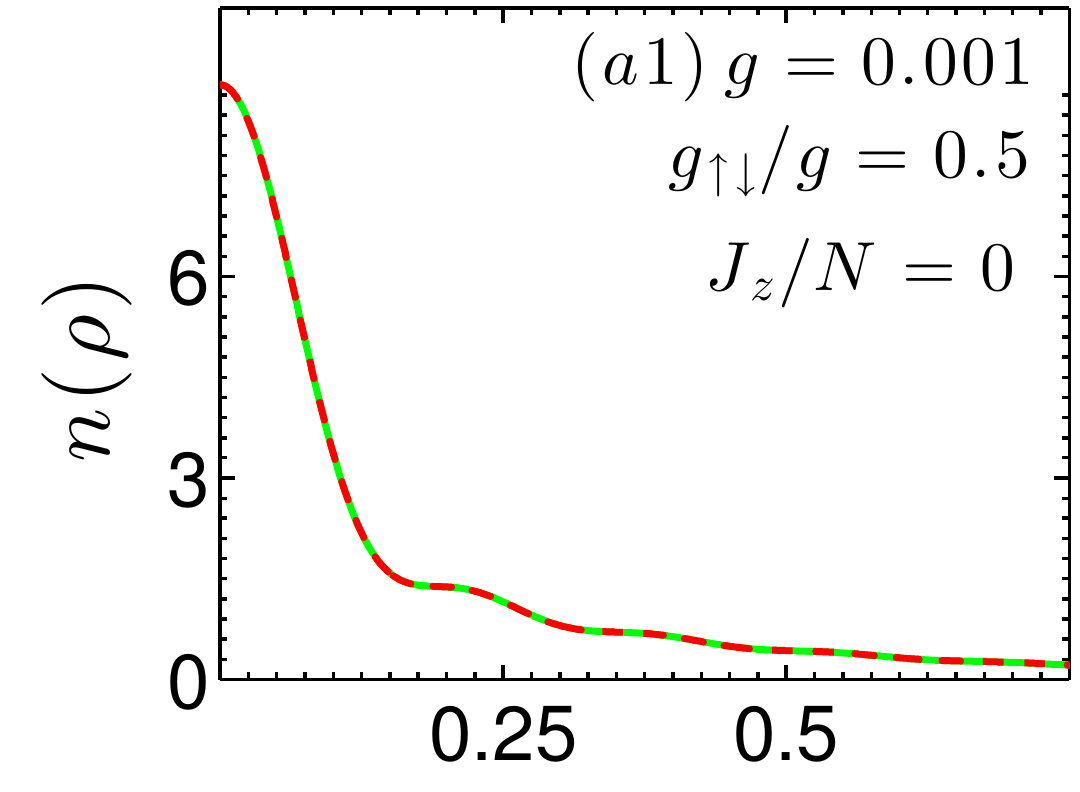} &
\includegraphics[clip,width=0.225\textwidth]{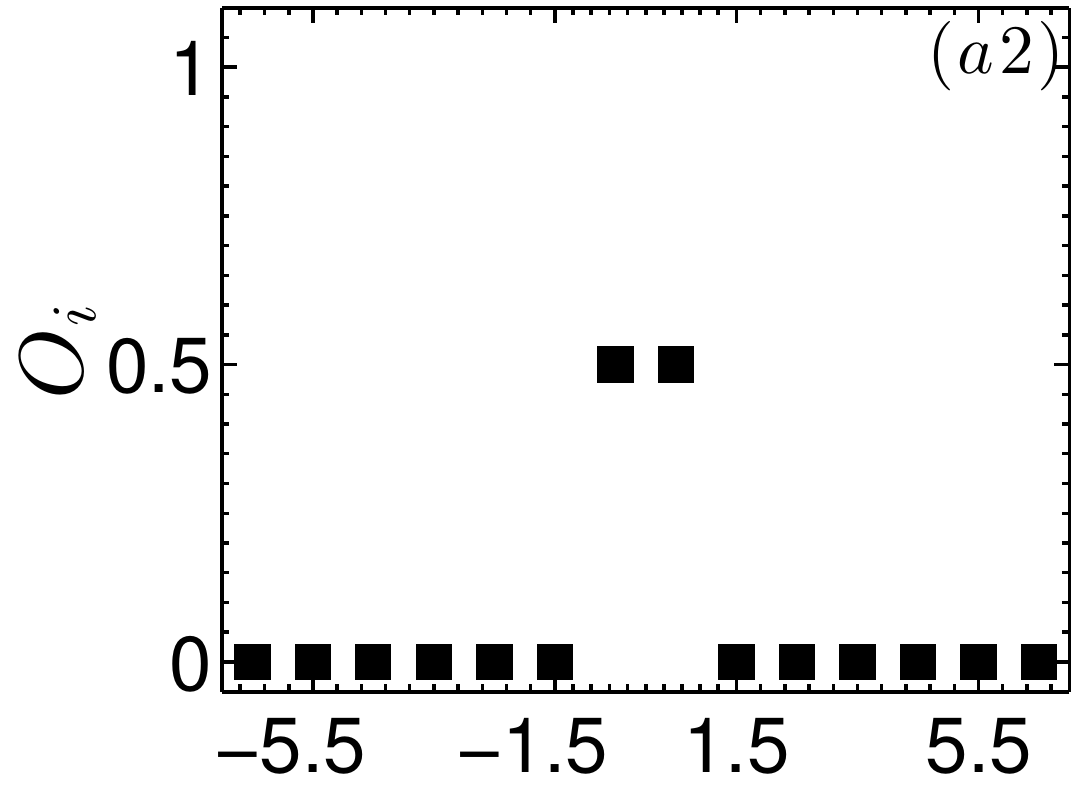} &
\includegraphics[clip,width=0.225\textwidth]{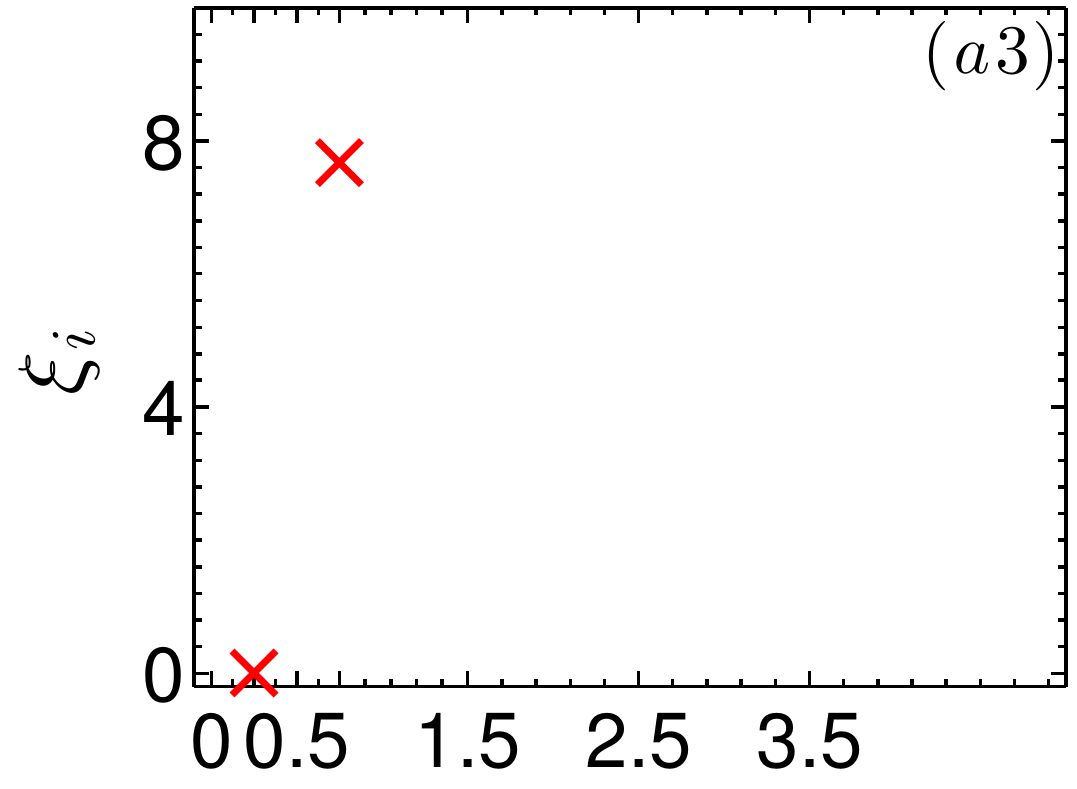} &
\includegraphics[clip,width=0.225\textwidth]{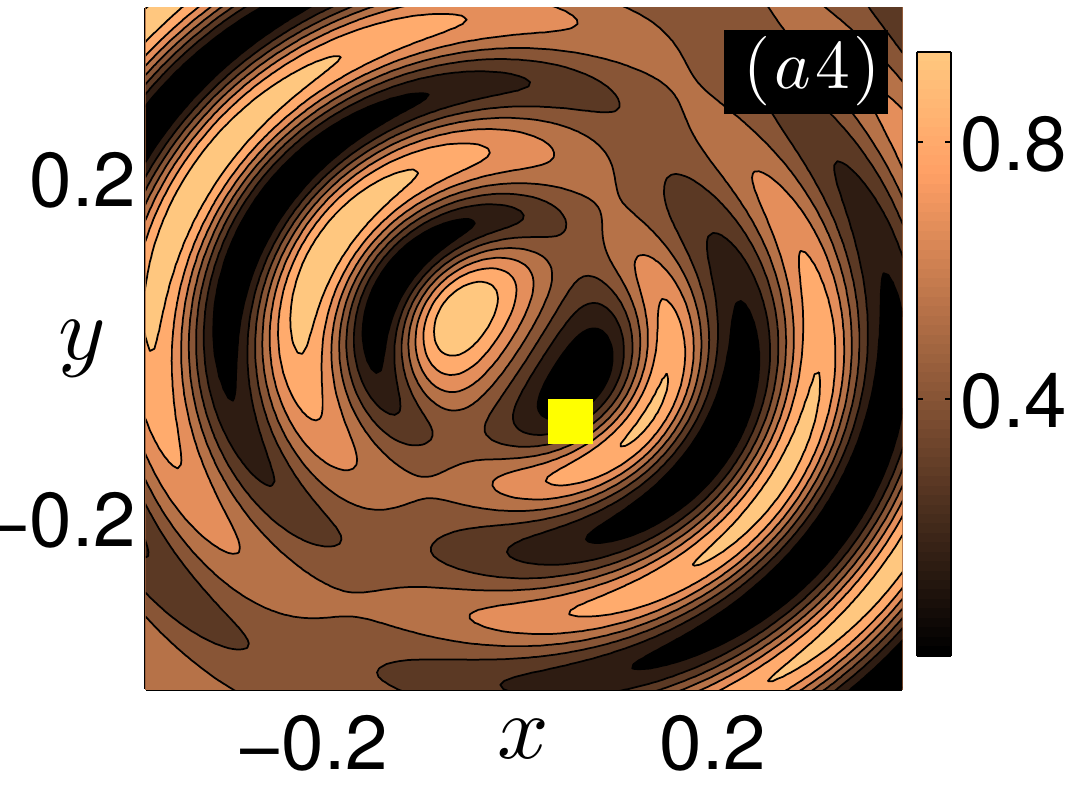} \\
\includegraphics[clip,width=0.225\textwidth]{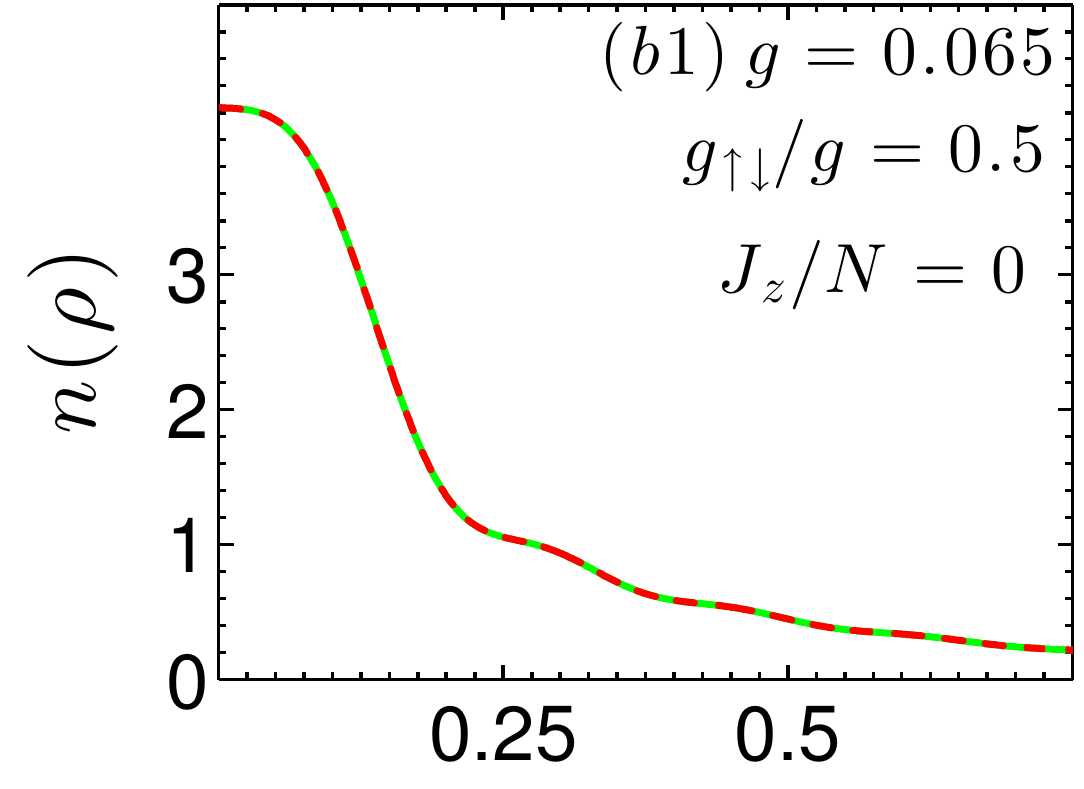} &
\includegraphics[clip,width=0.225\textwidth]{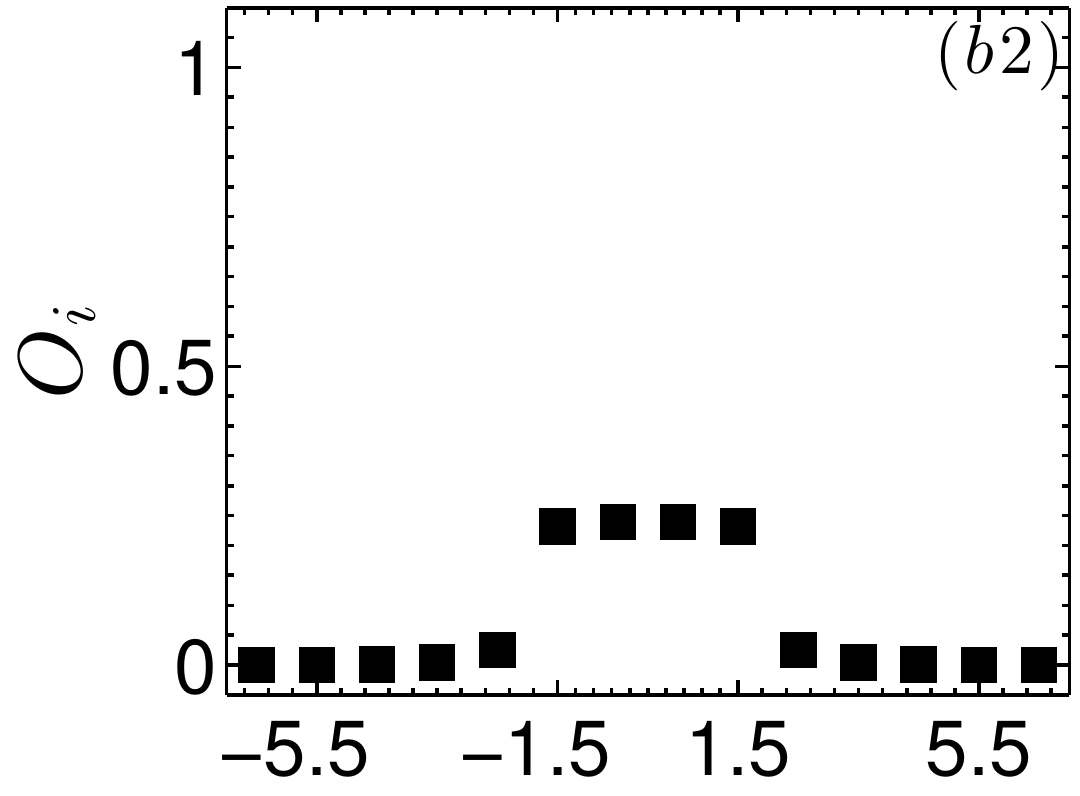} &
\includegraphics[clip,width=0.225\textwidth]{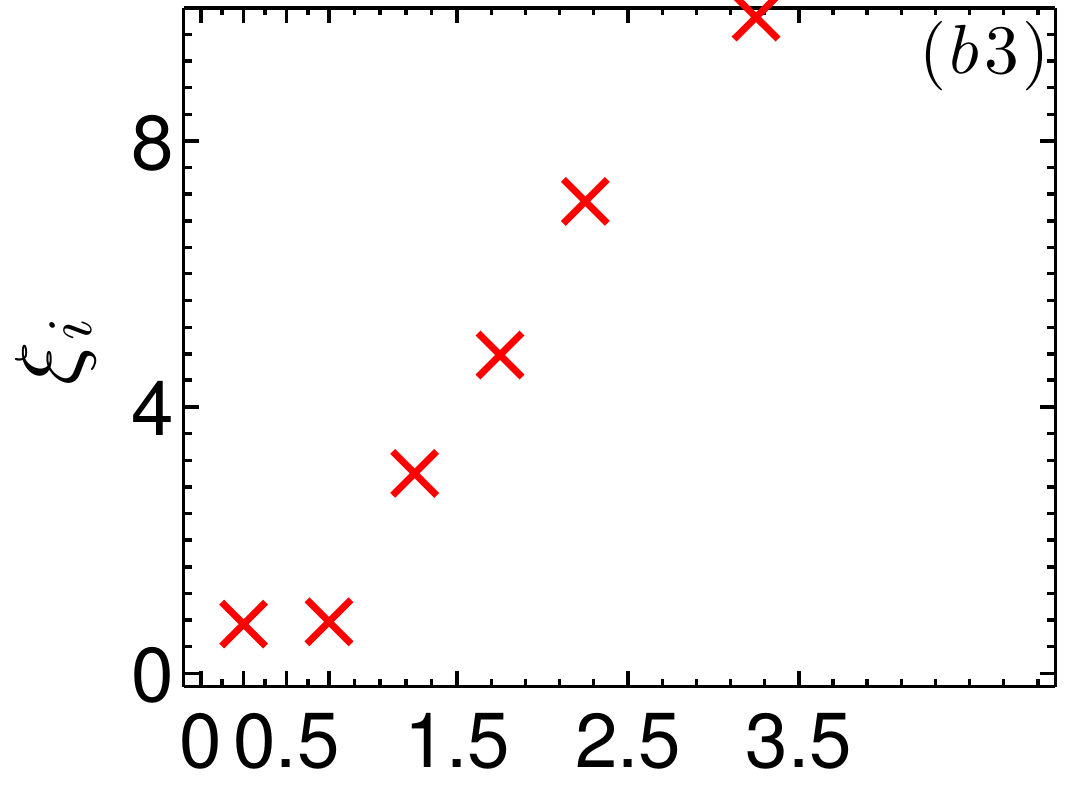} &
\includegraphics[clip,width=0.225\textwidth]{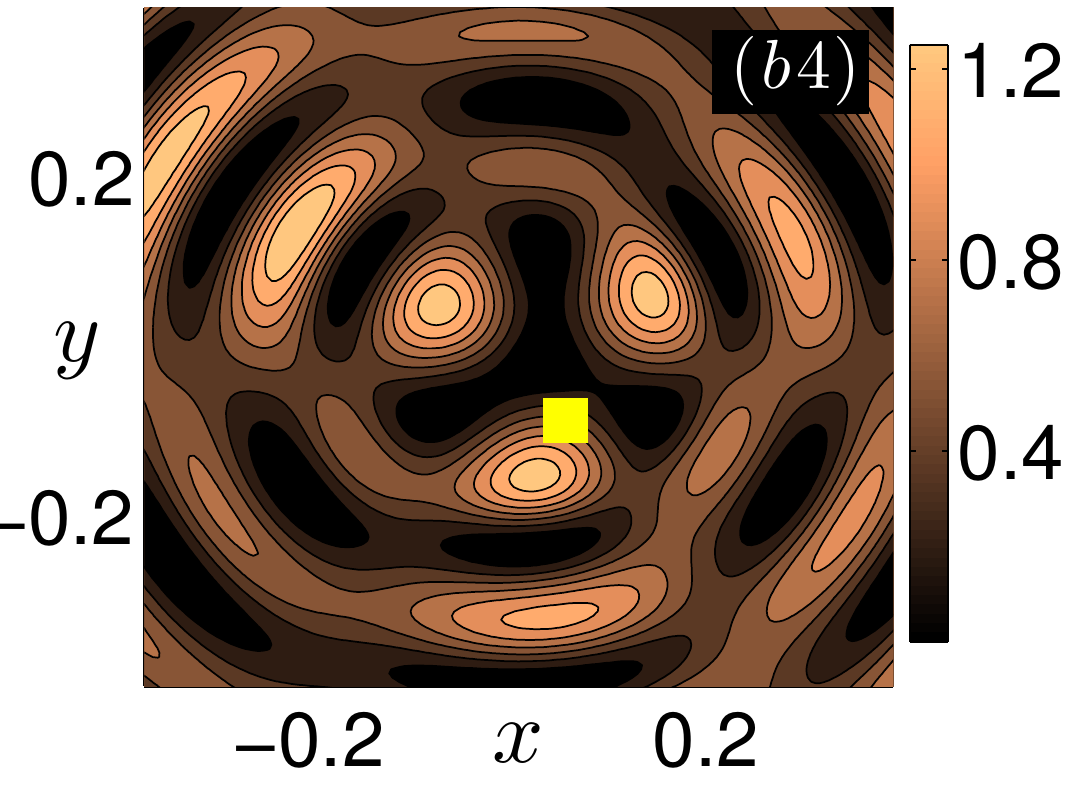} \\
\includegraphics[clip,width=0.225\textwidth]{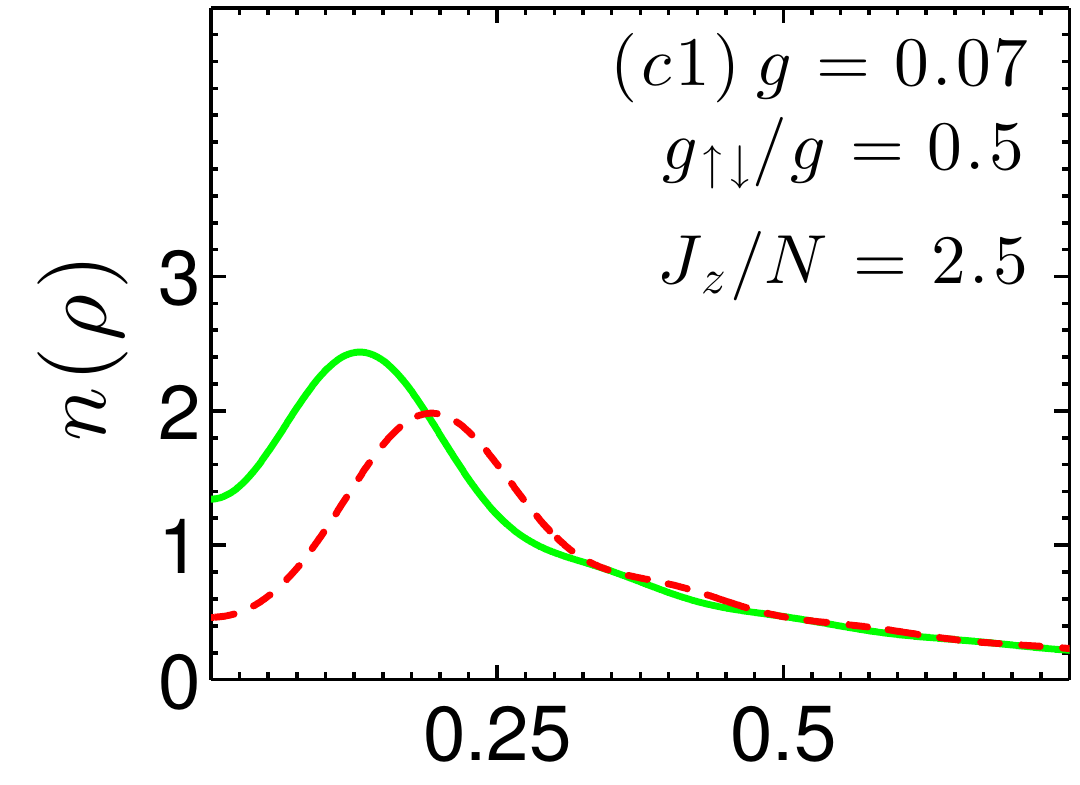} &
\includegraphics[clip,width=0.225\textwidth]{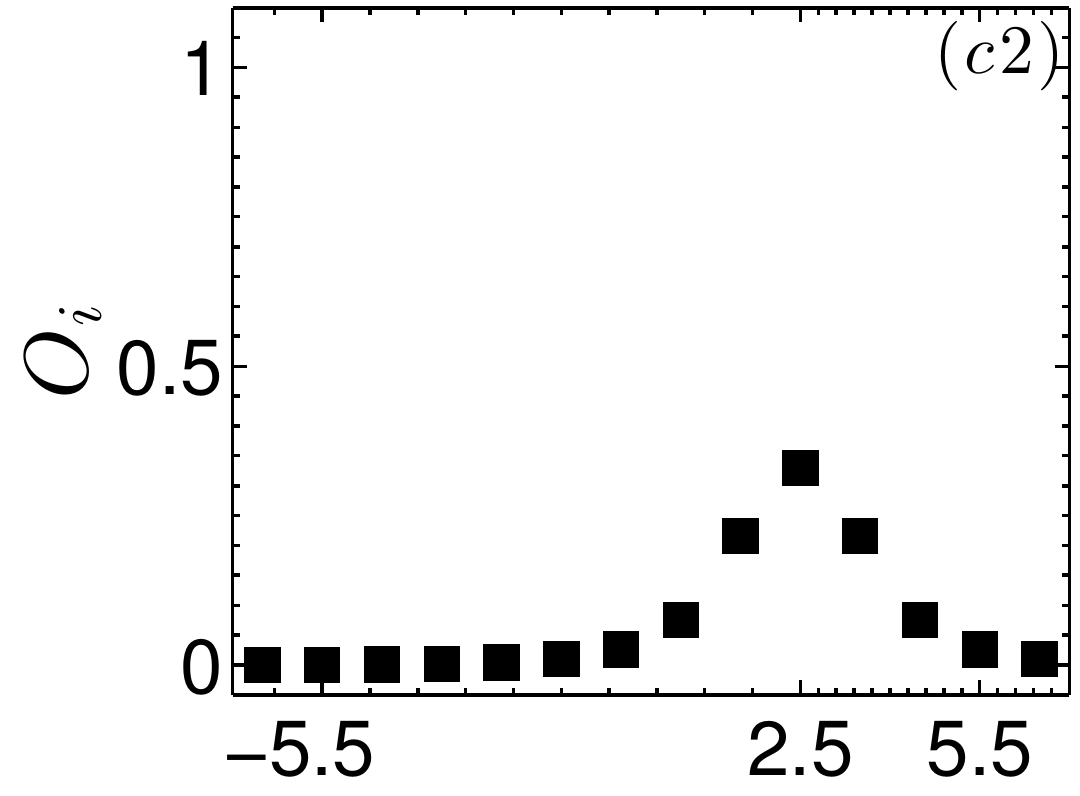} &
\includegraphics[clip,width=0.225\textwidth]{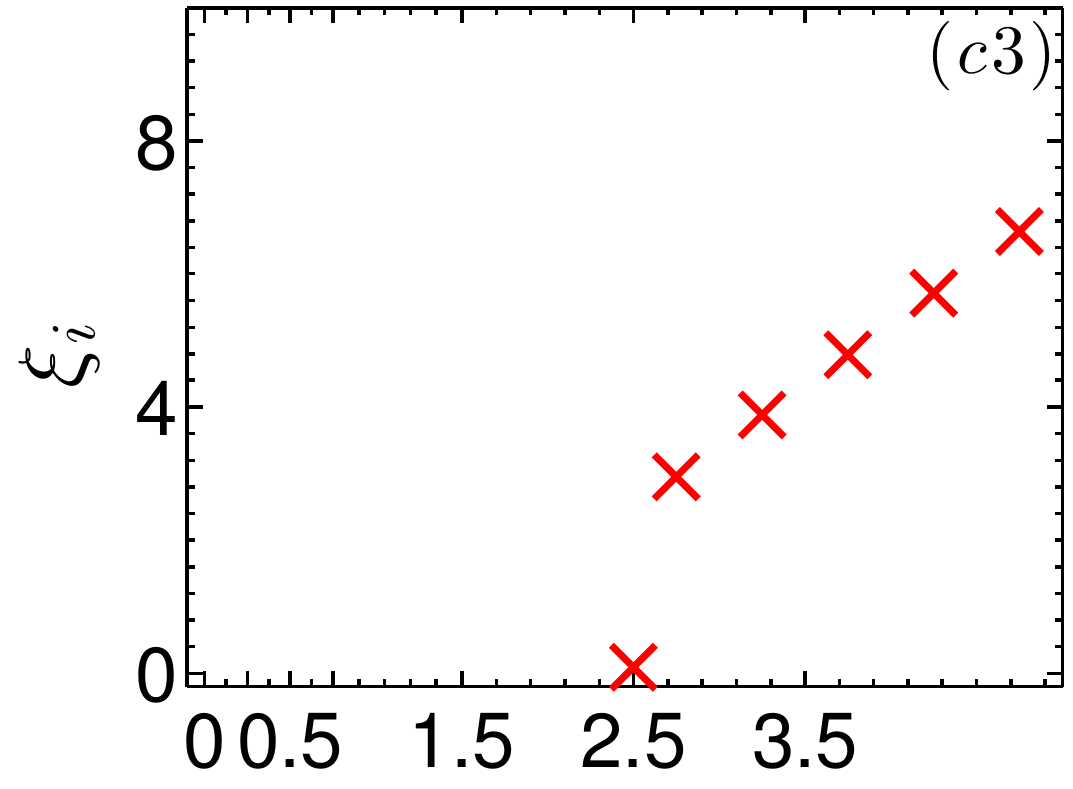} &
\includegraphics[clip,width=0.225\textwidth]{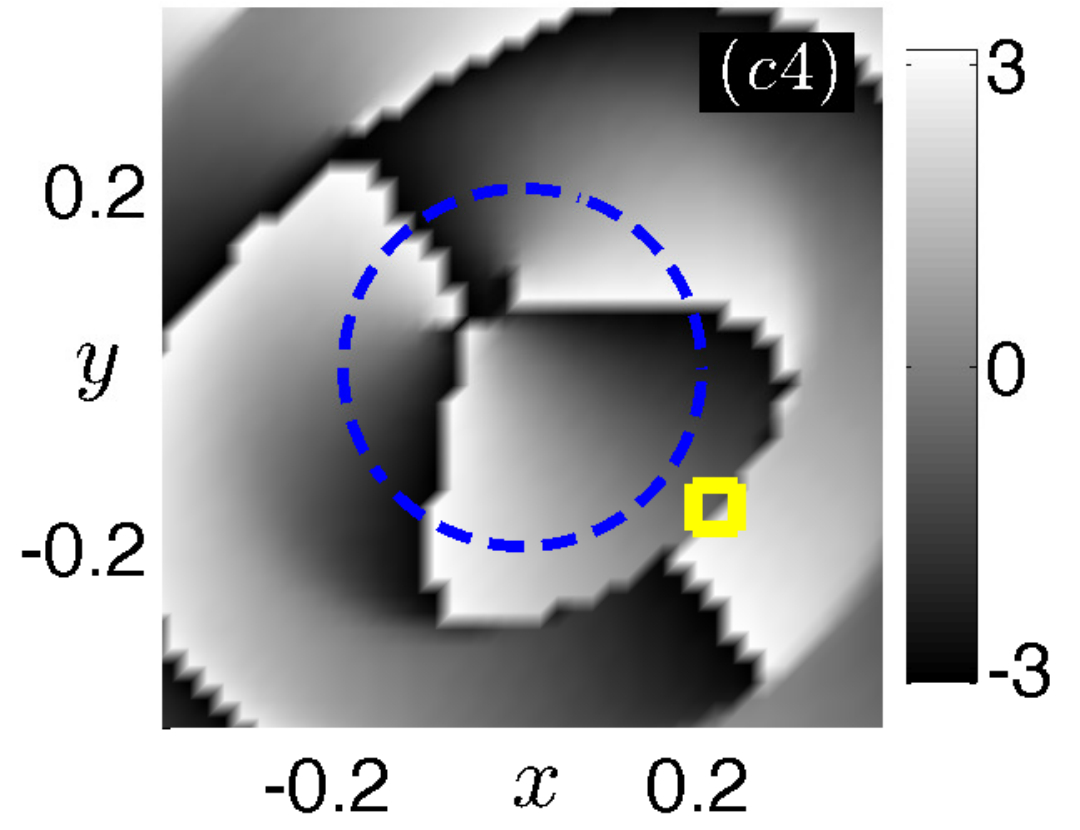} \\
\includegraphics[clip,width=0.225\textwidth]{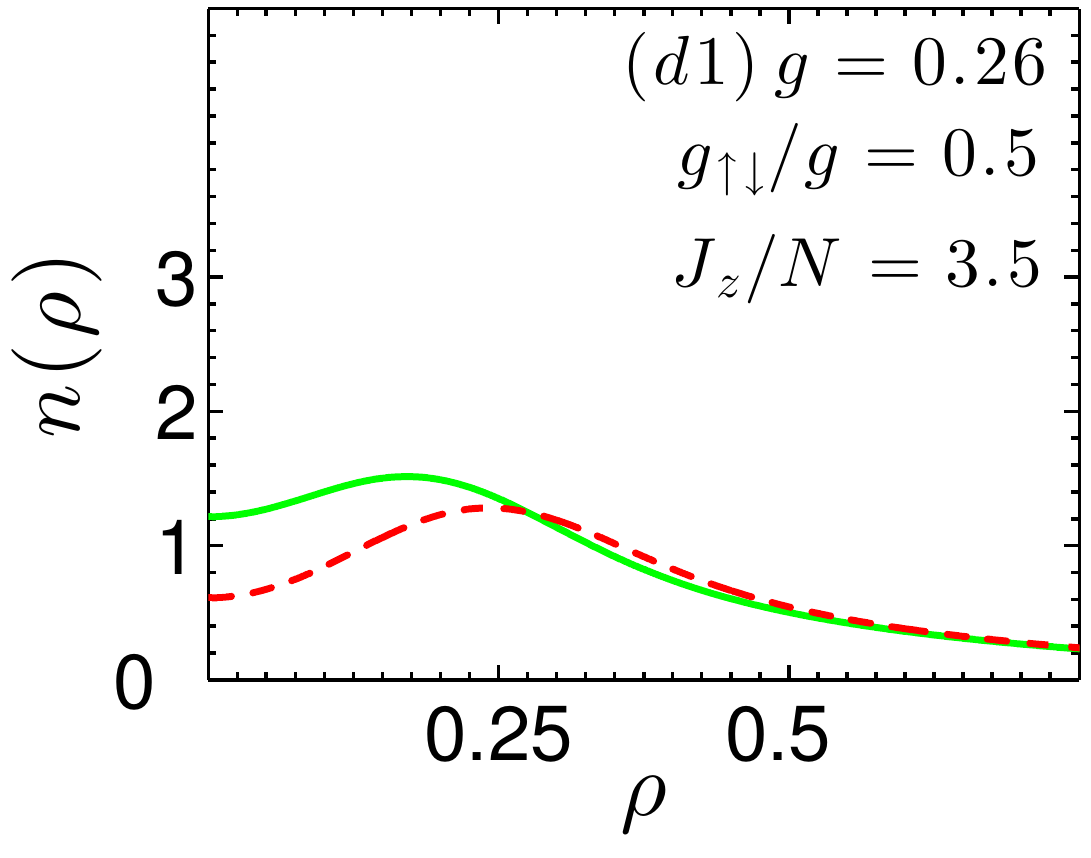} &
\includegraphics[clip,width=0.225\textwidth]{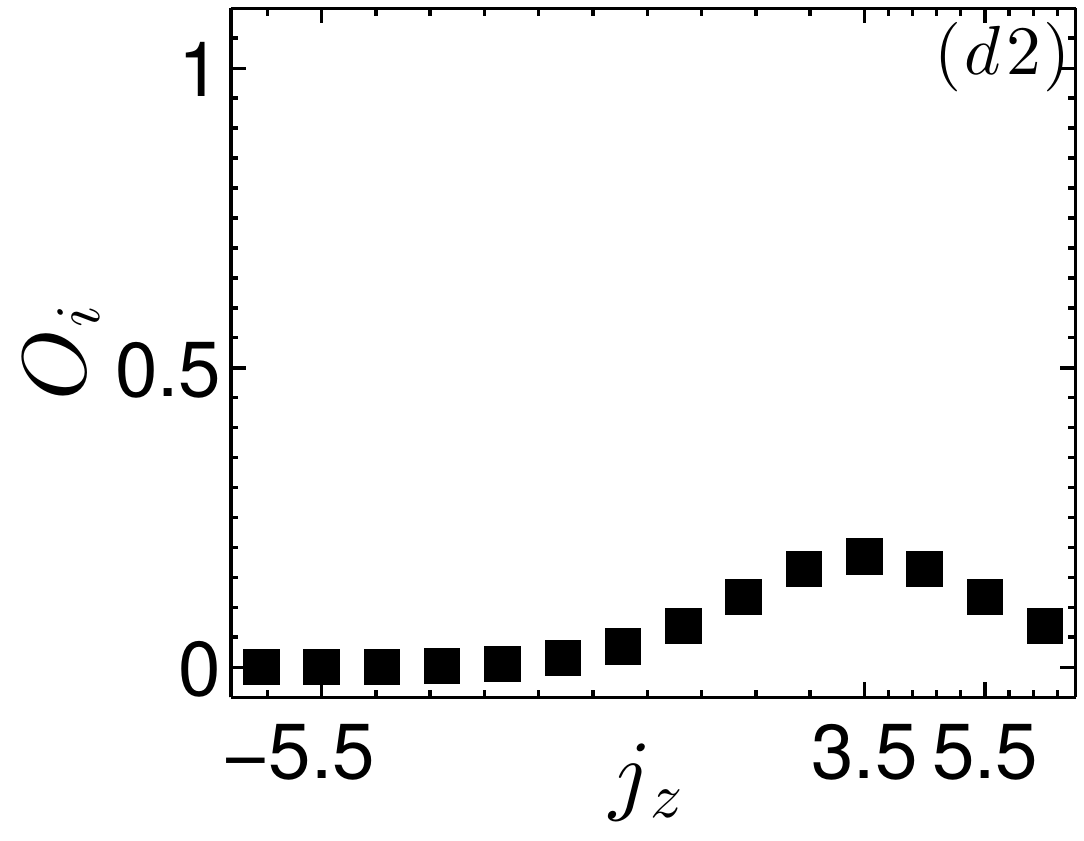} &
\includegraphics[clip,width=0.225\textwidth]{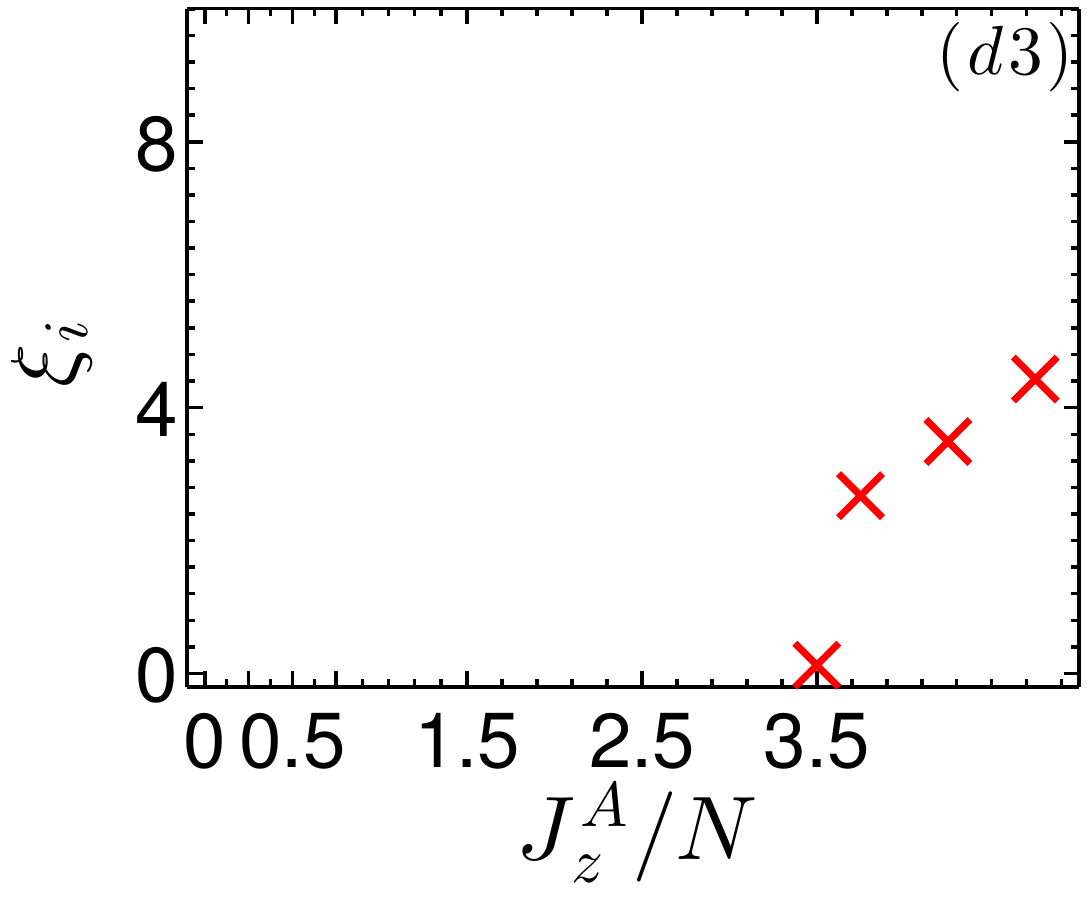} &
\includegraphics[clip,width=0.225\textwidth]{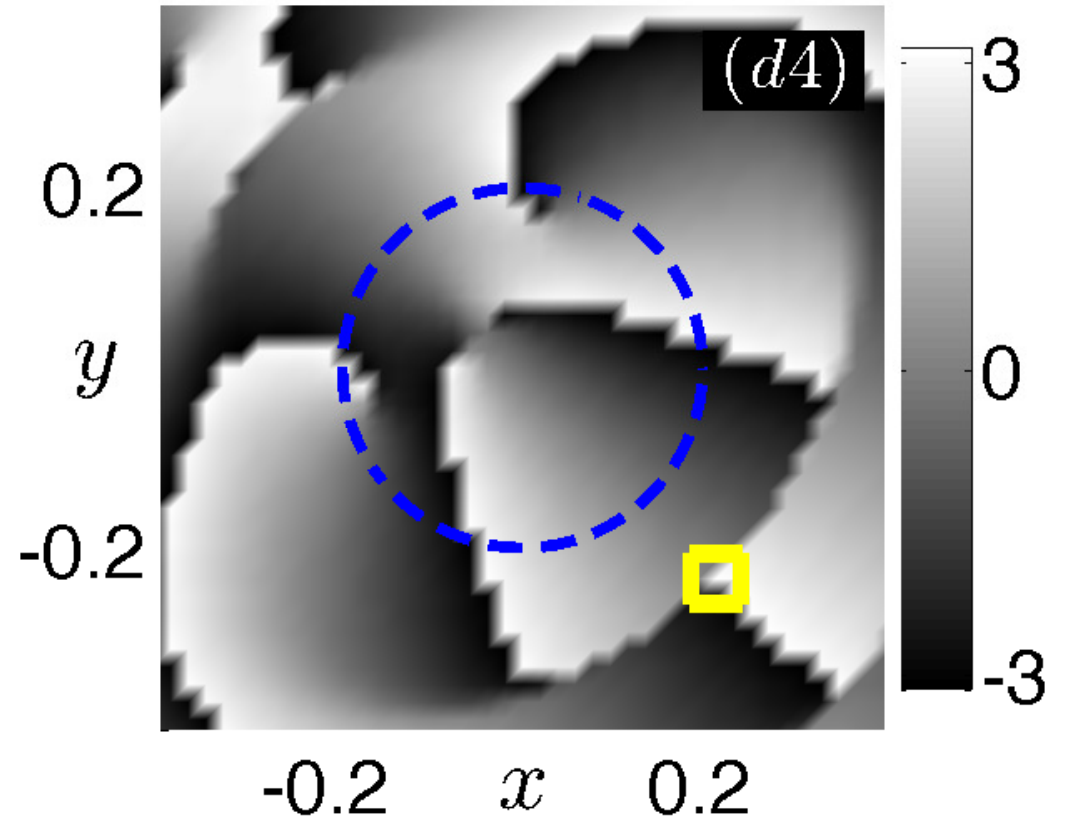} \\
\end{tabular}
\caption{(color online). Plots in each row illustrate the ground state properties at a representative interaction strength of Fig.~\ref{figN2ee05}$(a)$. In the first column (from left), we show density distributions of spin-up component $n_{\uparrow}(\rho)$ (solid green) and of spin-down component $n_{\downarrow}(\rho)$ (dashed red). In the second column, we show eigenvalues $O_i$ of single-particle density matrix as a function of angular momentum $j_z$ of the single-particle states $|\phi_i(\textbf{r})\rangle$. In the third column, we show corresponding $OES$ plots of entanglement pseudo-energies $\xi_i$ as a function of $J_z^A/N$, the average angular momentum of subsystem $A$. In the last column, we show contour plots $(a4)$ and $(b4)$ that are normalized pair-correlation functions $\langle n_{\uparrow}(\textbf{r}_0) n_{\downarrow}(\textbf{r})\rangle$, with $\textbf{r}_0$ denoted by a (yellow) marker. Phase plots $(c4)$ and $(d4)$ are derived from reduced wavefunction $\psi_{c,\downarrow}({\bf r})$, which is computed by fixing one of the two particles at their most probable locations and their corresponding radii are indicated by (yellow) markers. The closed dashed (blue) contour is a guide to the eye, that allows us to count the number of phase slips.}
\label{figN2all}
\end{figure*}
\par
\textbf{Figs.~\ref{figN2ee05}($\boldsymbol{a}$), \ref{figN2ee15}($\boldsymbol{a}$):} We solve the interacting Hamiltonian $\cal H$ at various interaction strengths and identify corresponding ground state manifolds $J_z/N$ in Figs.~\ref{figN2ee05}$(a)$ and \ref{figN2ee15}$(a)$. It is evident from the phase diagram that depending on $g$ and $g_{\uparrow\downarrow}$, the ground states belong to different $J_z/N$ manifolds. Furthermore, we determine if the ground state wavefunction   $\Psi_G$ is an eigenstate of ${\cal PT}$ operator, and thereby identify whether the state belongs to ${\cal P}$ or ${\cal PT}$ symmetry phase. In a broader sense, it is evident that a ground state in ${\cal PT}$ symmetry phase belongs to $J_z/N=0$ manifold, while ground states in various $J_z/N\neq0$ manifolds belong to ${\cal P}$ symmetry phase. $EE$ plots in Figs.~\ref{figN2ee05}$(b)$ and \ref{figN2ee15}$(b)$ reveal correlation properties in various phases. For pedagogical purposes, before we explain the features in $EE$ plots, we first discuss the symmetry, topological and correlation properties of ground states.
\par
In Fig.~\ref{figN2all}, we illustrate density distributions, eigenvalues of single-particle density matrix, orbital entanglement spectrum, pair-correlation functions and reduced wavefunctions at representative interaction strengths within various $J_z/N$ manifolds of Fig.~\ref{figN2ee05}$(a)$. Using a similar line of reasoning, we may understand the properties of ground states in Fig.~\ref{figN2ee05}$(b)$. Let us now proceed to discuss various plots shown in Fig.~\ref{figN2all}. 
\par
\textbf{Figs.~\ref{figN2all}($\boldsymbol{a1}$) $\rightarrow$ \ref{figN2all}($\boldsymbol{a4}$):} In this top row, we discuss the ground state properties of the ${\cal PT}$ eigenstate in $J_z/N=0$ manifold at $g=0.001$ of Fig.~\ref{figN2ee05}$(a)$. As shown in Fig.~\ref{figN2all}$(a1)$, the cylindrically symmetric density distributions $n_{\uparrow}(\rho)$ and $n_{\downarrow}(\rho)$ overlap. Being a ${\cal PT}$ eigenstate, it is evident from Fig.~\ref{figN2all}$(a2)$ that the positive and negative angular momentum states are equally occupied. Furthermore, the time-reversal partner states identified by quantum numbers $j_z = \pm 0.5$ are predominantly occupied. As expected, from the corresponding $OES$ plot in Fig.~\ref{figN2all}$(a3)$, we observe that the predominant contribution to the ground state is from the entanglement pseudo-energy $\xi_i$ at $J_z^A/N=+0.25$. From Figs.~\ref{figN2all}$(a2)$ and \ref{figN2all}$(a3)$, it is clear that the the maximally contributing Fock state is $\Phi_{{\cal PT}} = \mid n_{j_z=-0.5}=1, n_{j_z=+0.5}=1\rangle$, which explains the overlapping density distributions of  $n_{\uparrow}(\rho)$ and $n_{\downarrow}(\rho)$ in Fig.~\ref{figN2all}$(a1)$. In Fig.~\ref{figN2all}$(a4)$, we plot the (normalized) pair-correlation function $\langle n_{\uparrow}(\textbf{r}_0) n_{\downarrow}(\textbf{r})\rangle$ of this ${\cal PT}$ eigenstate. This plot illustrated the conditional probability to find a down-spin, when an up-spin component is assumed to be at a fixed point $\textbf{r}_0$, and reveals the presence of correlated regions (magnitude closer to 1) and anti-correlated regions (magnitude closer to 0). This plot illustrates the correlations present between up-spin and down-spin components that are not revealed by the cylindrically symmetric density distributions. 
\par
\textbf{Figs.~\ref{figN2all}($\boldsymbol{b1}$) $\rightarrow$ \ref{figN2all}($\boldsymbol{b4}$):} In this second row, we discuss the ground state properties of the ${\cal PT}$ eigenstate in $J_z/N=0$ manifold at $g=0.065$ of Fig.~\ref{figN2ee05}$(a)$. As discussed with reference to Fig.~\ref{figN2all}$(a1)$, the density distributions $n_{\uparrow}(\rho)$ and $n_{\downarrow}(\rho)$ overlap in Fig.~\ref{figN2all}$(b1)$. It is evident from Fig.~\ref{figN2all}$(b2)$ that the time-reversal partner states identified by $j_z = \pm 0.5$ and $j_z = \pm 1.5$ are almost equally occupied. From the corresponding $OES$ plot in Fig.~\ref{figN2all}$(b3)$, we observe that the ground state is equally occupied by $\xi_i$ at $J_z^A/N=+0.25$ and $+0.75$. From Figs.~\ref{figN2all}$(b2)$ and \ref{figN2all}$(b3)$, it is clear that the the maximally contributing Fock states are $\Phi_{{\cal PT}} = \mid n_{j_z=-0.5}=1, n_{j_z=+0.5}=1\rangle$ and $\Phi_{{\cal PT}} = \mid n_{j_z=-1.5}=1, n_{j_z=+1.5}=1\rangle$. To illustrate the internal structure of this ${\cal PT}$ eigenstate and the correlations between up-spin and down-spin components, we show the pair-correlation function in Fig.~\ref{figN2all}$(b4)$. 
\par
\textbf{Figs.~\ref{figN2all}($\boldsymbol{c1}$) $\rightarrow$ \ref{figN2all}($\boldsymbol{c4}$):} In this third row, we discuss the ground state properties of the ${\cal P}$ eigenstate in $J_z/N=+2.5$ manifold at $g=0.07$ of Fig.~\ref{figN2ee05}$(a)$. While the corresponding ground state is degenerate in $J_z/N= \pm2.5$ manifolds, we restrict our discussion to  $J_z/N= +2.5$ manifold without loss of generality. The cylindrically symmetric density distributions $n_{\uparrow}(\rho)$ and $n_{\downarrow}(\rho)$ are distinct, as shown in Fig.~\ref{figN2all}$(c1)$. In this ${\cal P}$ eigenstate, there is an inherent asymmetry in the occupation of positive and negative angular momentum states. This is evident from the plot of single-particle density matrix eigenvalues $O_i$ in Fig.~\ref{figN2all}$(c2)$. This explains the presence of distinct density distributions in Fig.~\ref{figN2all}$(c1)$. Furthermore, we observe a peak in the occupation of eigenstate identified by $j_z = + 2.5$ in Fig.~\ref{figN2all}$(c2)$. From the corresponding $OES$ plot in Fig.~\ref{figN2all}$(c3)$, we observe that the ground state is predominantly occupied by $\xi_i$ at $J_z^A/N=2.5$. To illustrate the internal structure of this ${\cal P}$ eigenstate, we show the phase plot derived from the reduced wavefunction $\psi_{c,\downarrow}({\bf r})$ in Fig.~\ref{figN2all}$(c4)$. To better understand this phase plot, we take cues from plots in Figs.~\ref{figN2all}$(c2)$ and \ref{figN2all}$(c3)$. Though we observe from Fig.~\ref{figN2all}$(c3)$ that the ground state is predominantly occupied by $\xi_i$ at $J_z^A/N=2.5$, it may be conceived from Fig.~\ref{figN2all}$(c2)$ that the ground state has contributions from various Fock states, for example: $\Phi_{\cal P} = \mid n_{j_z=+2.5}=2\rangle$ or $\Phi_{\cal P} = \mid n_{j_z=+1.5}=1, n_{j_z=+3.5}=1\rangle$ or $\Phi_{\cal P} = \mid n_{j_z=+0.5}=1, n_{j_z=+4.5}=1\rangle$. From the representation of single-particle eigenstates in Eqn.~(\ref{eqnspstates}), it is evident that the net orbital angular momentum of spin-up component in the ground state is +2 and that of spin-down component is +3. Correspondingly, the phase plot of the down-spin component in Fig.~\ref{figN2all}$(c4)$ reveals a vorticity of 3. We note here that the vorticity is the number of phase slips from $+\pi$ to $-\pi$, i.e., when the shadowing changes from white to black. For convenience, we identify this ${\cal P}$ eigenstate as ${\cal P}3$, where 3 is the vorticity of the down-spin component. 
\par
\textbf{Figs.~\ref{figN2all}($\boldsymbol{d1}$) $\rightarrow$ \ref{figN2all}($\boldsymbol{d4}$):} In this last row, we discuss the ground state properties of the ${\cal P}$ eigenstate in $J_z/N=+3.5$ manifold at $g=0.26$ of Fig.~\ref{figN2ee05}$(a)$. The corresponding ground state is degenerate in $J_z/N= \pm3.5$ manifolds, while we restrict our discussion to $J_z/N= +3.5$ manifold. As expected for a ${\cal P}$ eigenstate, the density distributions $n_{\uparrow}(\rho)$ and $n_{\downarrow}(\rho)$ shown in Fig.~\ref{figN2all}$(d1)$ are distinct. In addition to the asymmetric occupation of positive and negative angular momentum states in Fig.~\ref{figN2all}$(d2)$, we observe a peak occupation of eigenstate identified by $j_z = + 3.5$. From the corresponding $OES$ plot in Fig.~\ref{figN2all}$(d3)$, we observe that the ground state is predominantly occupied by $\xi_i$ at $J_z^A/N=3.5$. However, it may be conceived from Fig.~\ref{figN2all}$(d2)$ that the ground state has contributions from various Fock states, for example: $\Phi_{\cal P} = \mid n_{j_z=+3.5}=2\rangle$ or $\Phi_{\cal P} = \mid n_{j_z=+1.5}=1, n_{j_z=+5.5}=1\rangle$ or $\Phi_{\cal P} = \mid n_{j_z=+2.5}=1, n_{j_z=+4.5}=1\rangle$. It is clear that with increasing inter-particle interaction strengths, the particles distribute themselves in higher angular momentum manifolds. Furthermore, it is evident that the net orbital angular momentum of spin-up component in the ground state is +3 and that of spin-down component is +4. Correspondingly, the phase plot of down-spin component in Fig.~\ref{figN2all}$(d4)$ reveals a vorticity of 4. For convenience, we identify this ${\cal P}$ eigenstate as ${\cal P}4$. We further note that the phase plots of down-spin components derived for ${\cal P}1$ and ${\cal P}2$ eigenstates in Fig.~\ref{figN2ee15}$(a)$ exhibit a vorticity of 1 and 2 respectively.
\par
\textbf{Figs.~\ref{figN2ee05}($\boldsymbol{b}$), \ref{figN2ee15}($\boldsymbol{b}$):} As noted earlier, $OES$ preserves the whole spectrum of eigenvalues of the RDM, and hence allows us to extract information about the occupation of Fock states with different subsystem angular momenta $J_z^A$. With our understanding of $OES$ plots in Figs.~\ref{figN2all}, we now proceed to explain various features observed in $EE$ plots of Figs.~\ref{figN2ee05}$(b)$ and \ref{figN2ee15}$(b)$. (\textbf{i}) The presence of distinctly different slopes suggests the presence of distinct correlation properties in ground states within various phases. (\textbf{ii}) Within each phase, $EE$ increases monotonously with increasing $g$. As discussed in Sec.~\ref{secem}, this results from an increasingly homogeneous distribution of Fock states in the ground state $J_z/N$ manifold, and in-turn an increased correlation. For example, to illustrate this feature within the ${\cal PT}$ symmetric phase in Fig.~\ref{figN2ee05}$(b)$, we may compare $OES$ plots in Fig.~\ref{figN2all}$(a3)$ and \ref{figN2all}$(b3)$ and observe an increased homogeneity in distribution of Fock states. (\textbf{iii}) The presence of nearly degenerate $\xi_i$ values results in a reduction in the \emph{slope} of $EE$. While this feature is observed at larger interaction strengths within the ${\cal PT}$ symmetric phase of Fig.~\ref{figN2ee05}$(b)$, the $OES$ plot in Fig.~\ref{figN2all}$(b3)$ helps us understand this. (\textbf{iv}) Transition to a ${\cal P}$ symmetric phase is marked by a sharp reduction in the \emph{value} of $EE$  \cite{comments1}. To better understand this feature, we compare $OES$ plots in Fig.~\ref{figN2all}$(b3)$ and \ref{figN2all}$(c3)$ and observe a sharp reduction in homogeneity of $\xi_i$ values, accompanied by a substantial drop in the minimum value of $\xi_i$. In summary, we emphasize that the knowledge of $OES$ helps us understand  various features exhibited by $EE$ plots.
\par
In summary, it is evident that the interacting Hamiltonian picks either a ${\cal P}$-eigenstate or a ${\cal PT}$-eigenstate to be the ground state. The ground state is sensitive to the relative magnitudes of $g_{\uparrow\downarrow}$ and $g$. $J_z/N$ plots allow us to identify various ${\cal P}$ and ${\cal PT}$ symmetry phases in the interacting system. With the analysis of density distributions, single-particle density matrix and reduced wavefunctions, we illustrate  ground state symmetry and topological properties. We assert that the bosons condense into an array of ${\cal P}$-symmetric topological ground states that have $n+1/2$ -quantum angular momentum vortex configuration, with $n = 0, 1, 2, 3$. With the analysis of single-particle density matrix, $OES$ and pair-correlation functions, we illustrate the internal structure of different ground states in the ${\cal PT}$ symmetry phase. We analyze the correlation properties of the ground states with the help of $OES$ and $EE$ plots. 
\begin{figure}[t*]
\begin{centering}
\includegraphics[clip,width=0.48\textwidth]{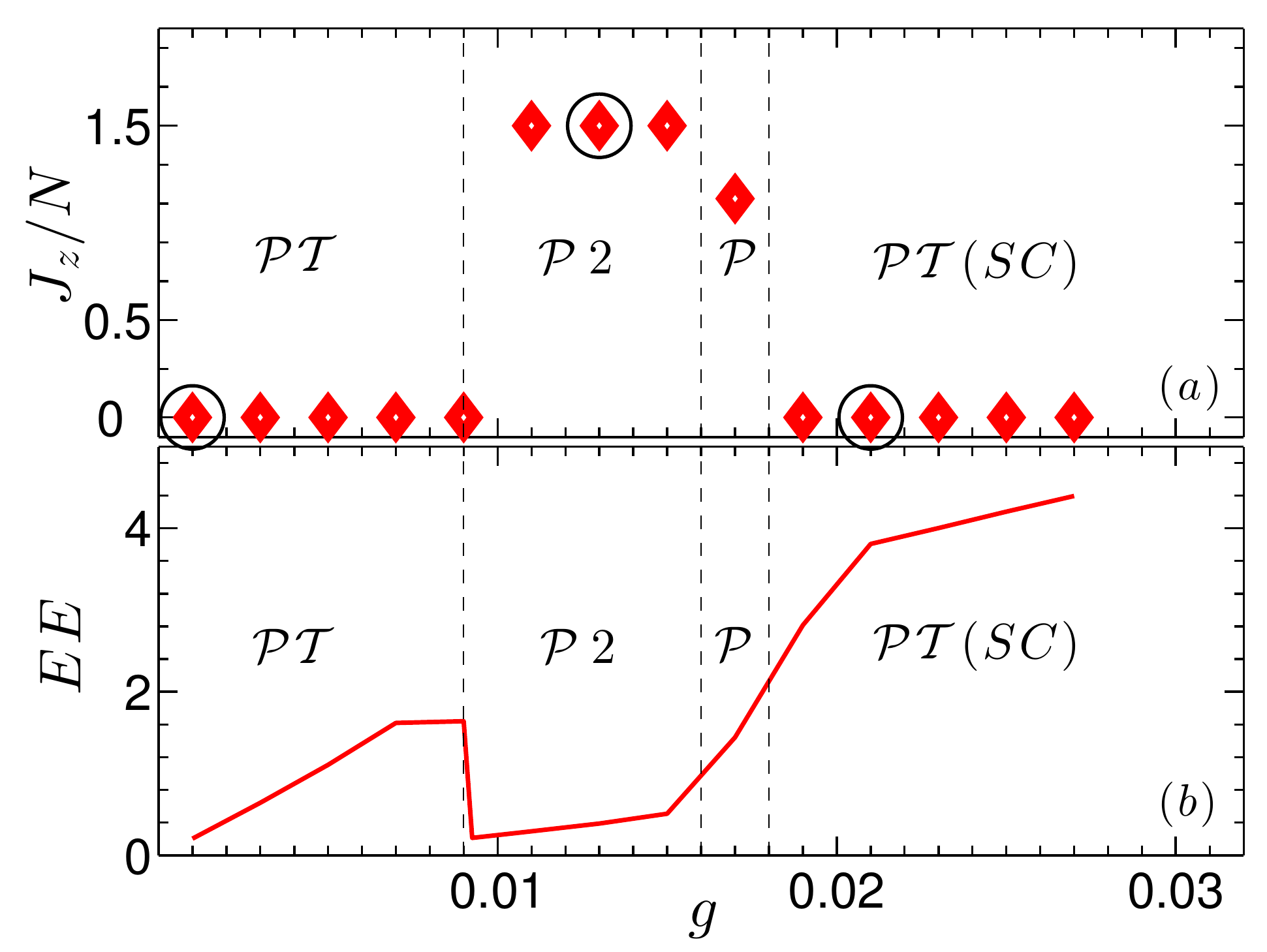} 
\par\end{centering}
\caption{(color online). Plots of $(a)$ ground state $J_z/N$ manifolds and $(b)$ entanglement entropy, as a function of interaction strength $g$ with $\lambda_{SO}=20, N = 8, g_{\uparrow\downarrow}/g=0.5$. For representative interaction strengths denoted by circled (black) markers, we illustrate the ground state properties in Fig.~\ref{figN8all}.}
\label{figN8ee05} 
\end{figure}
\begin{figure}[t*]
\begin{centering}
\includegraphics[clip,width=0.48\textwidth]{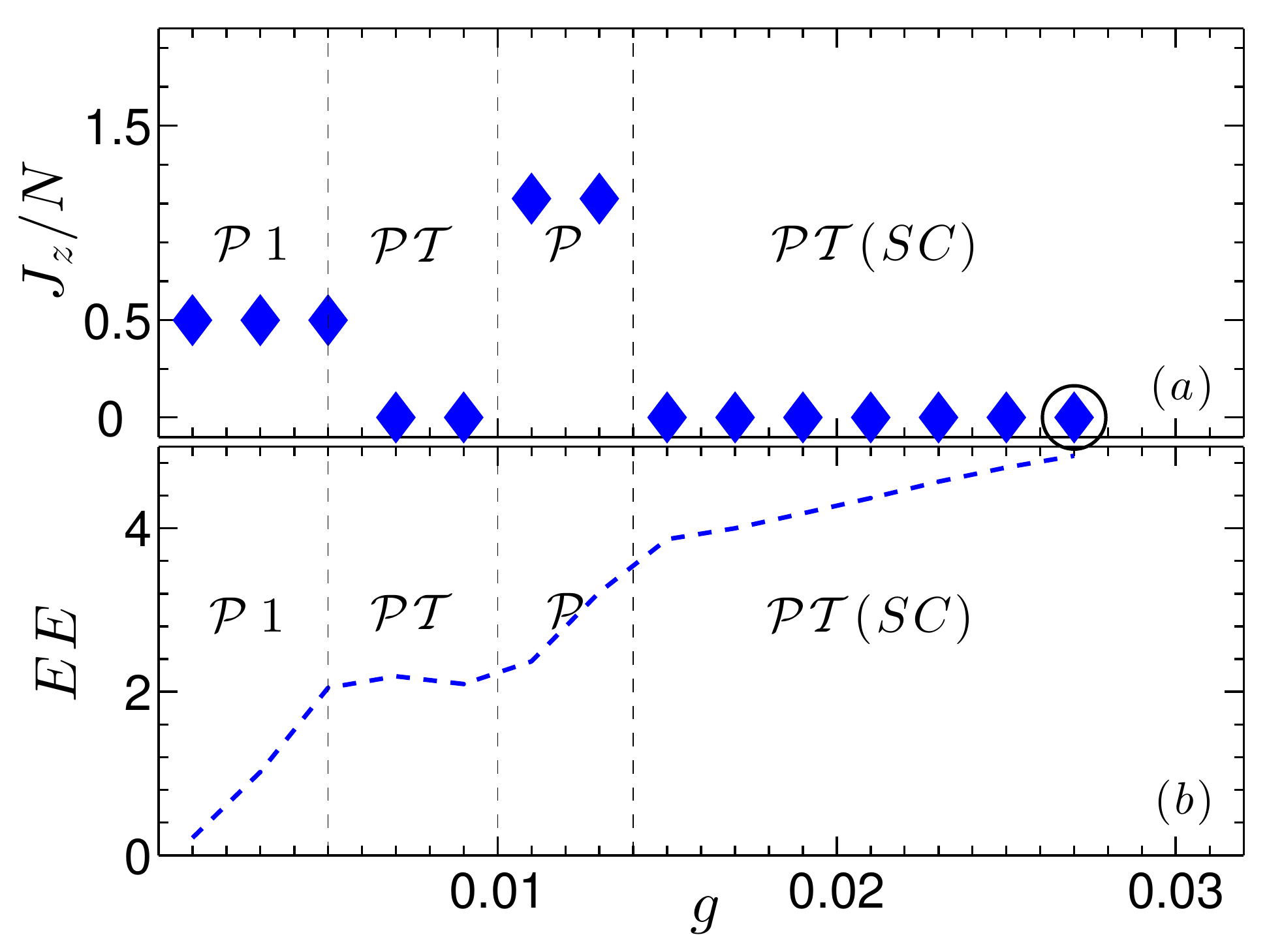} 
\par\end{centering}
\caption{(color online). Plots of $(a)$ ground state $J_z/N$ manifolds and $(b)$ entanglement entropy, as a function of interaction strength $g$ with $\lambda_{SO}=20, N = 8, g_{\uparrow\downarrow}/g=1.5$. For the representative interaction strength denoted by a circled (black) marker, we illustrate the ground state properties in Fig.~\ref{figN8all}.}
\label{figN8ee15} 
\end{figure}
\subsection{$N=8$}
\label{secn8}
\begin{figure*}[t!]
\centering
\begin{tabular}{cccc}
\includegraphics[clip,width=0.225\textwidth]{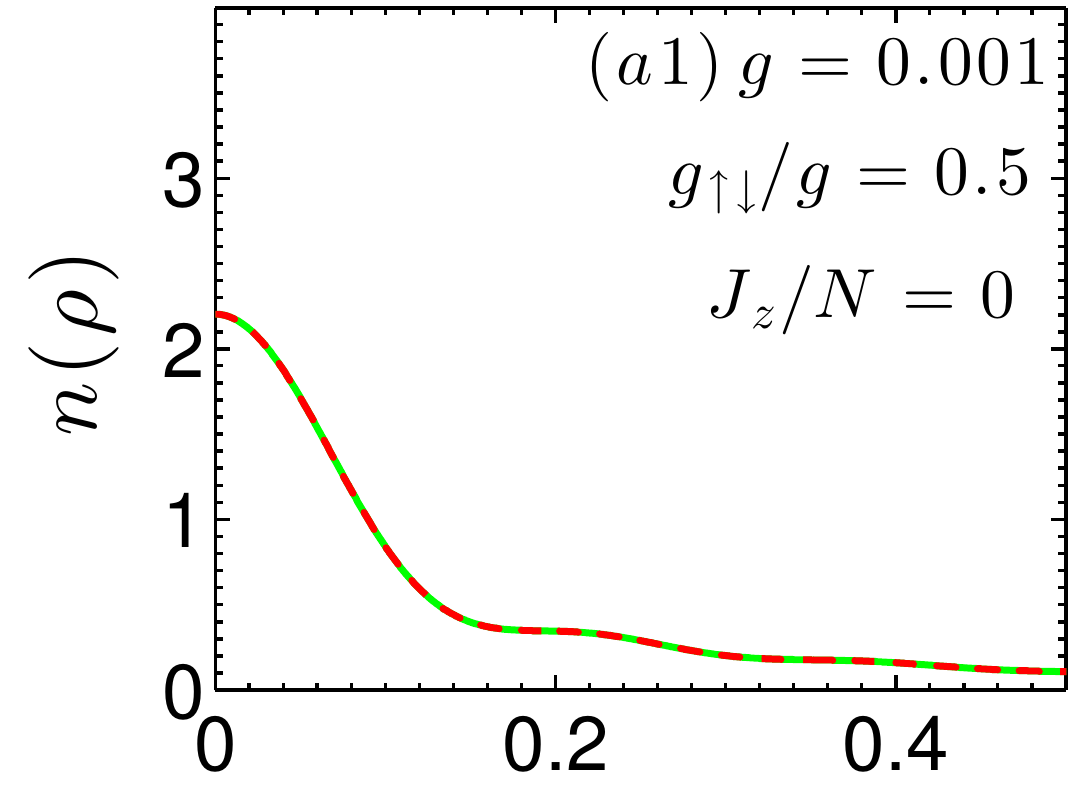} &
\includegraphics[clip,width=0.225\textwidth]{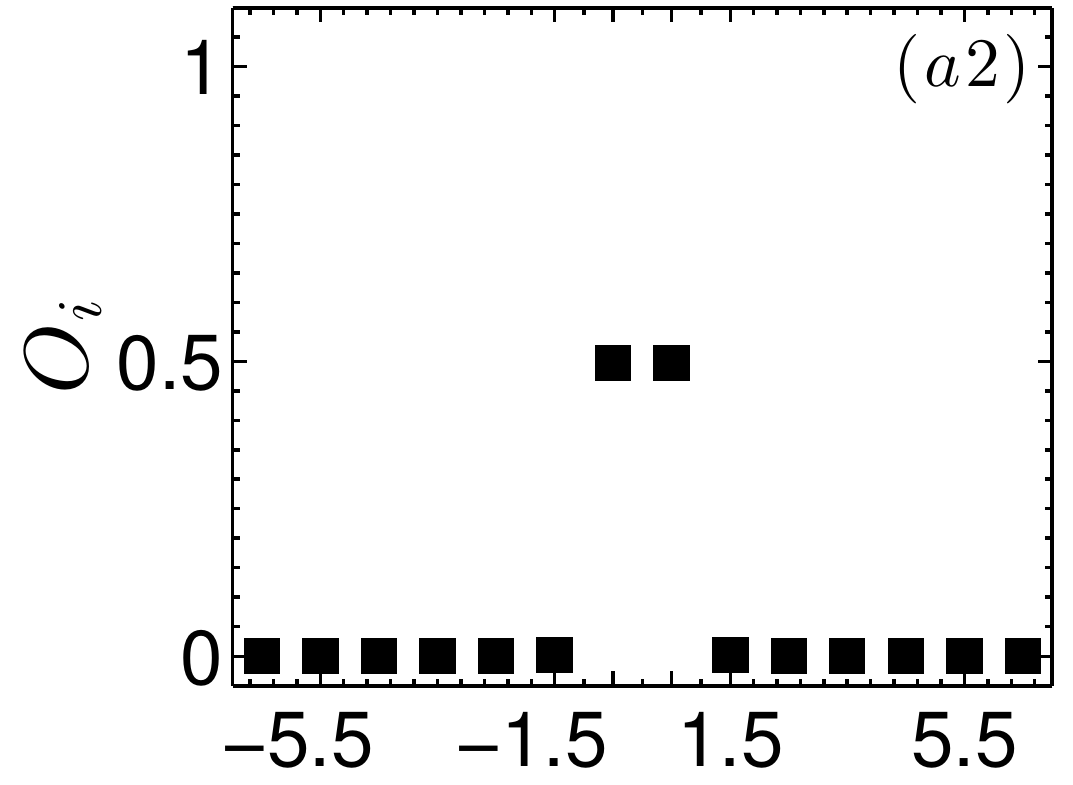} &
\includegraphics[clip,width=0.225\textwidth]{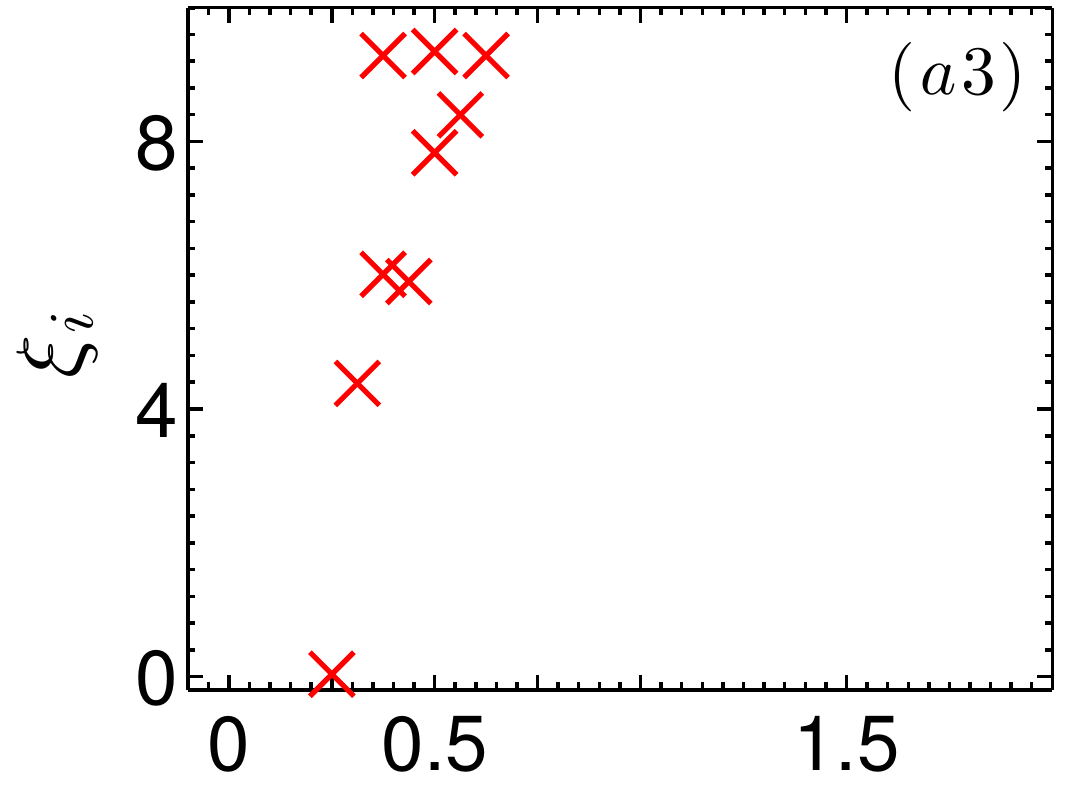} &
\includegraphics[clip,width=0.225\textwidth]{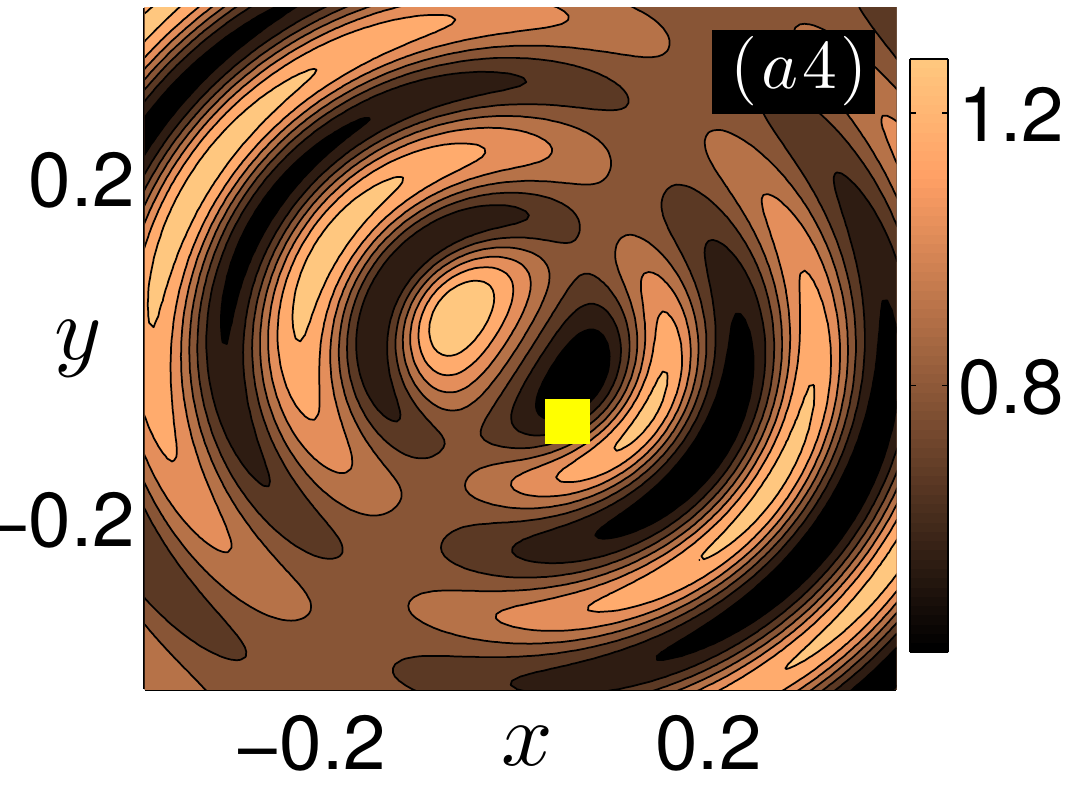} \\
\includegraphics[clip,width=0.225\textwidth]{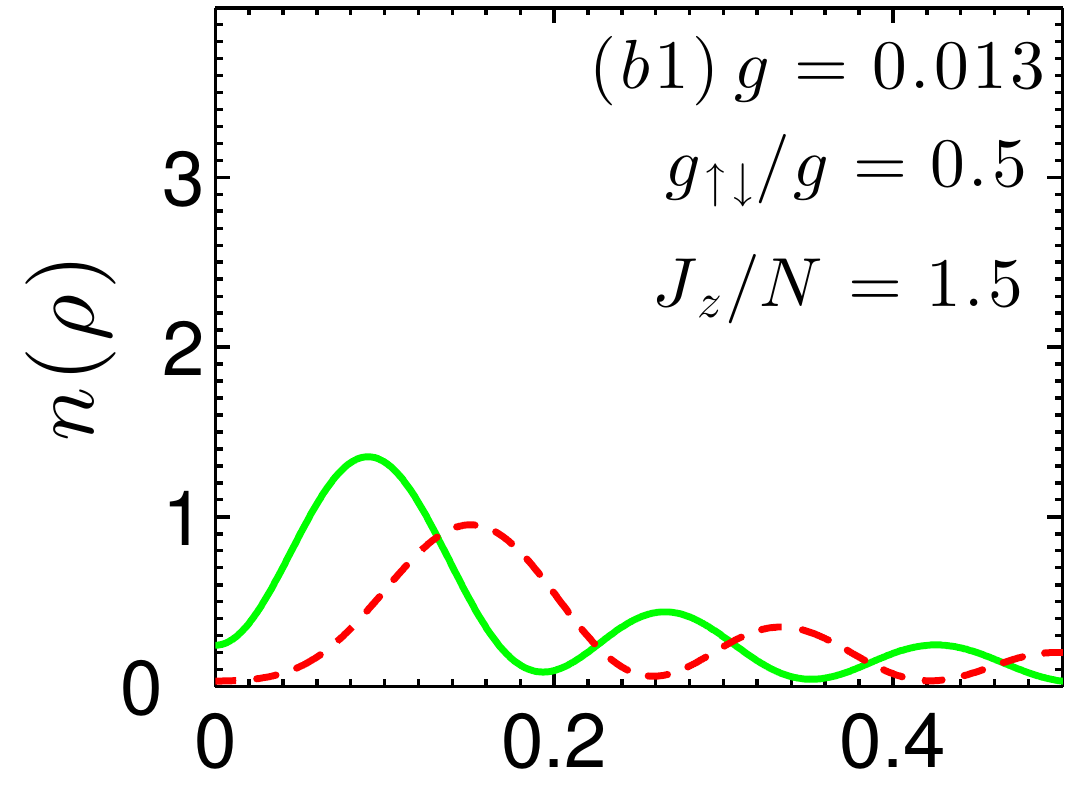} &
\includegraphics[clip,width=0.225\textwidth]{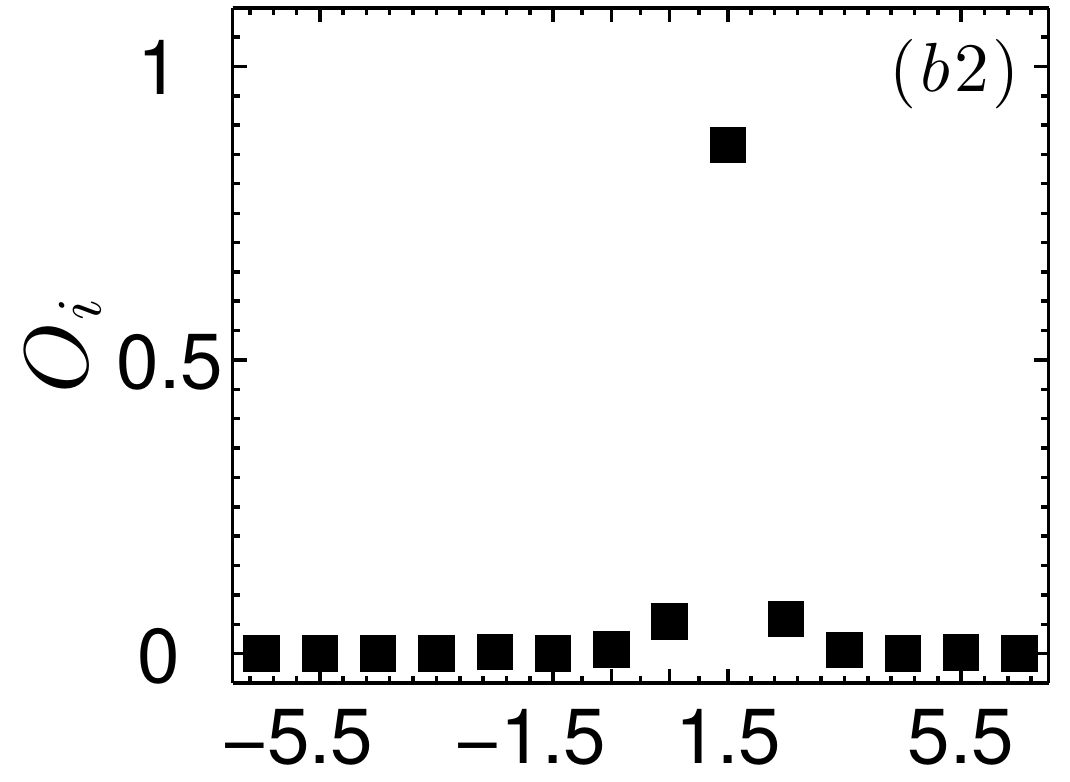} &
\includegraphics[clip,width=0.225\textwidth]{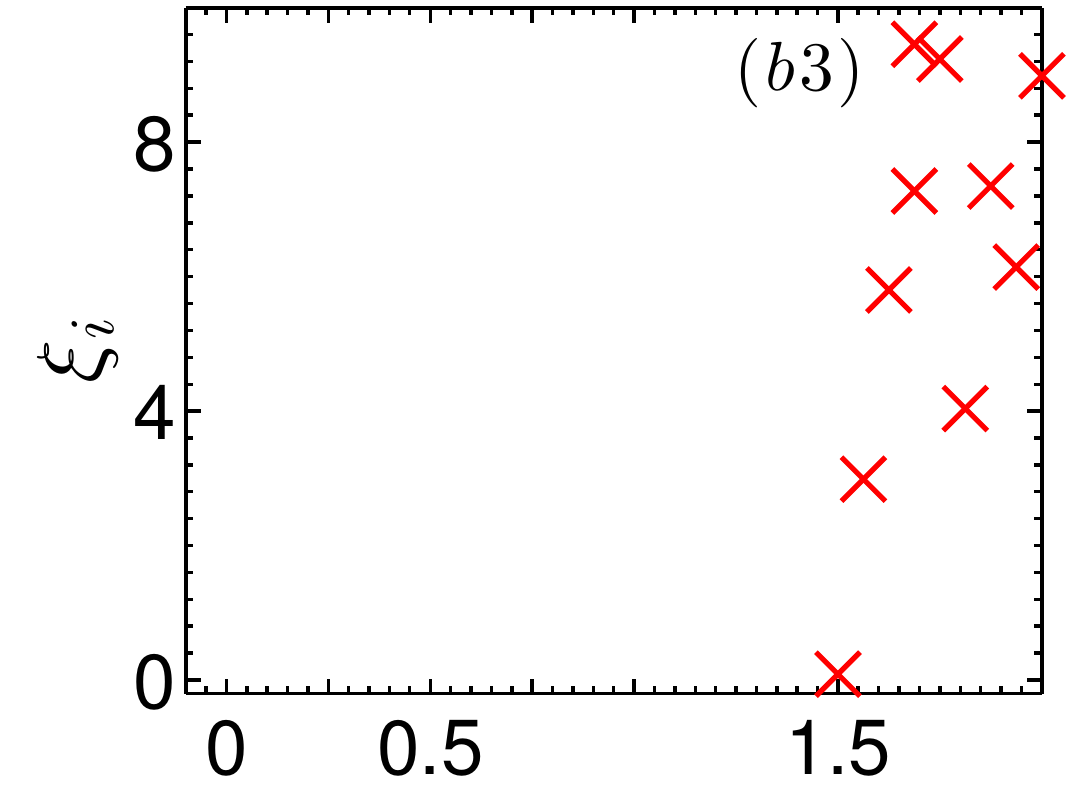} &
\includegraphics[clip,width=0.225\textwidth]{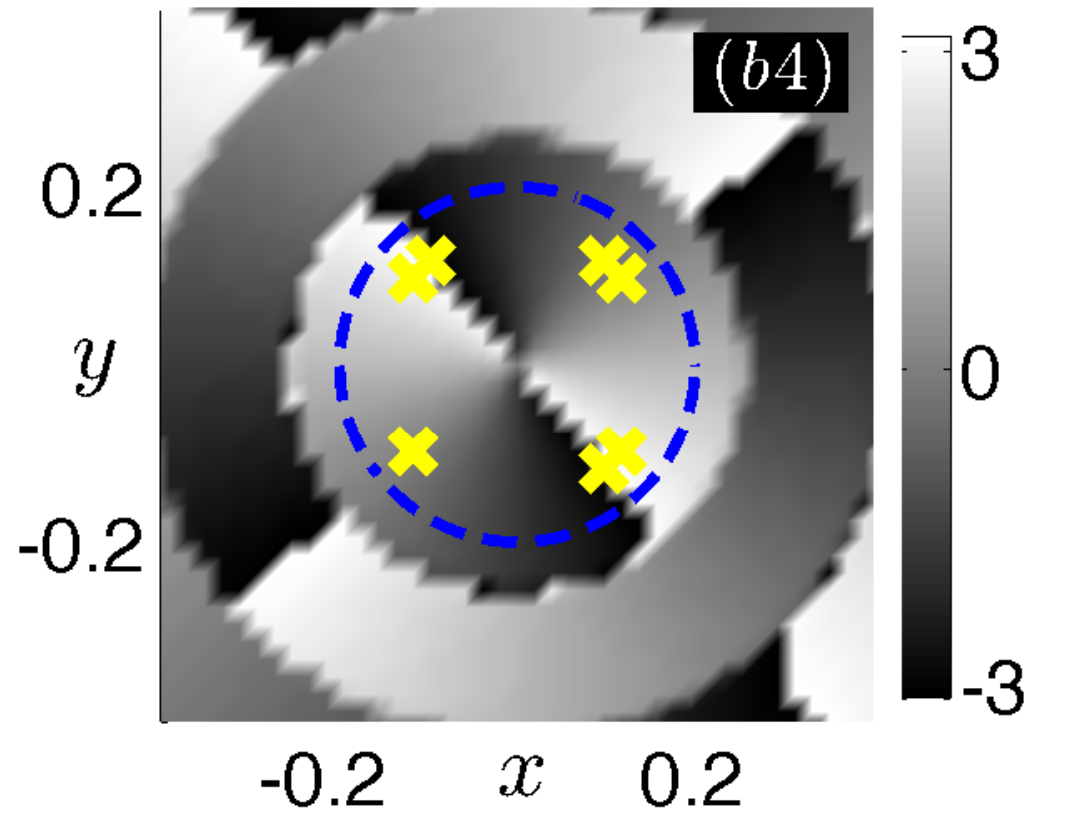} \\
\includegraphics[clip,width=0.225\textwidth]{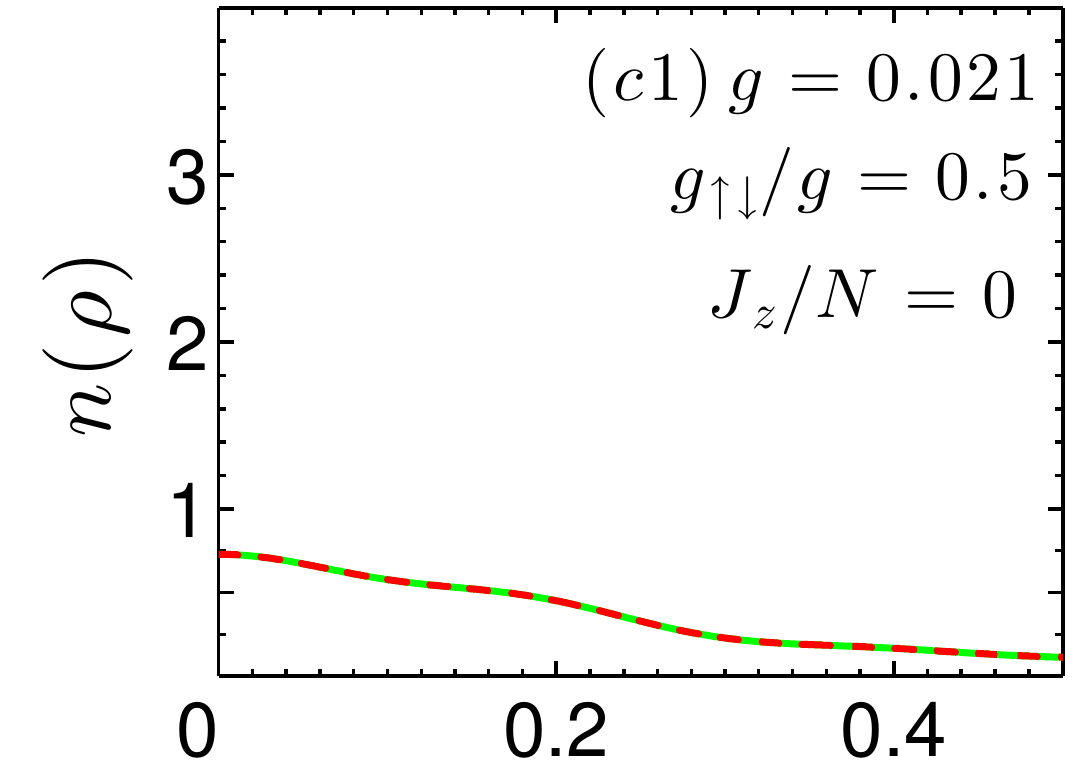} &
\includegraphics[clip,width=0.225\textwidth]{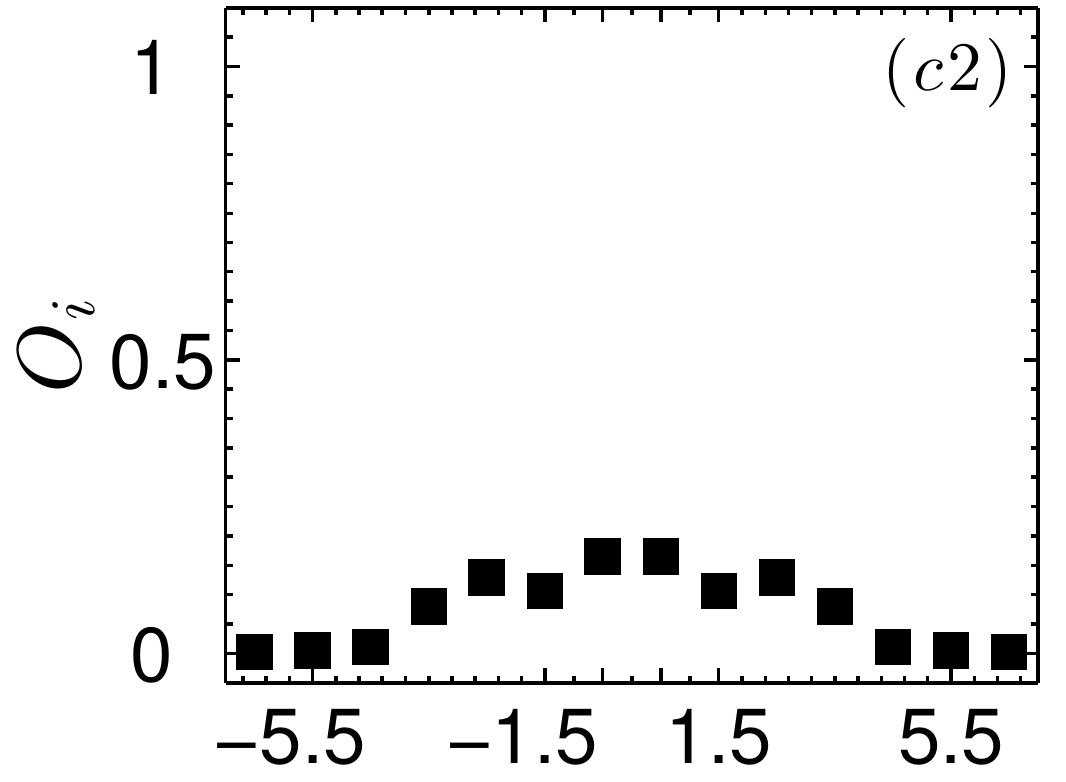} &
\includegraphics[clip,width=0.225\textwidth]{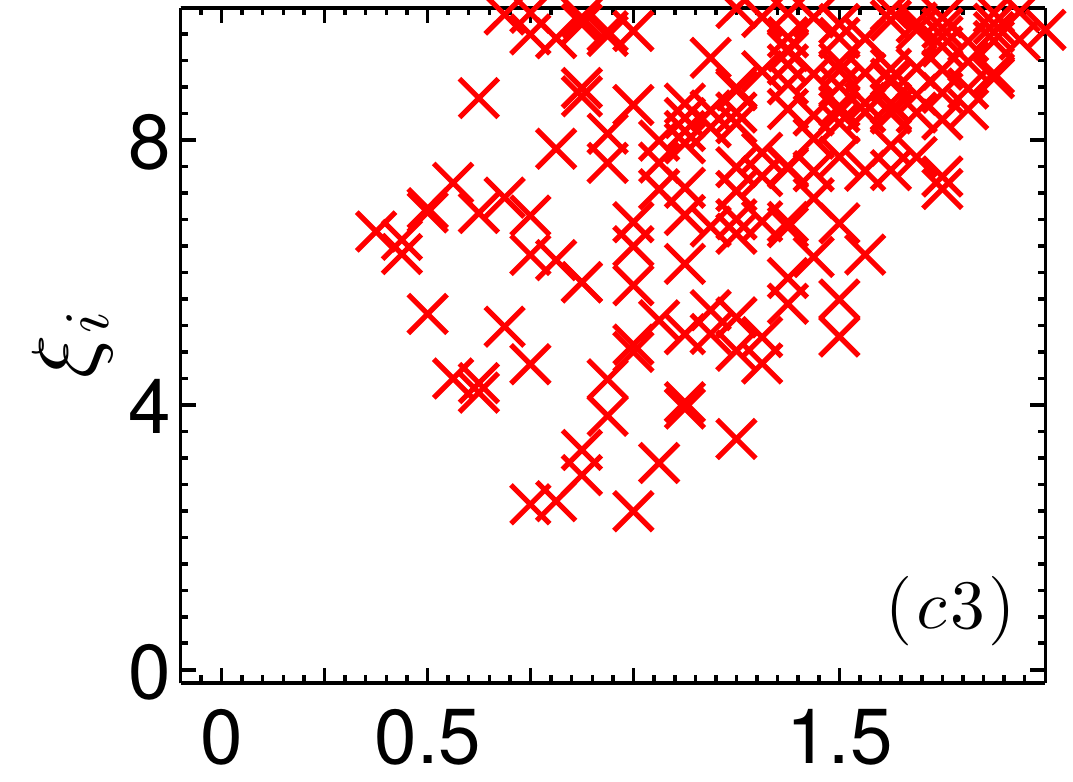} &
\includegraphics[clip,width=0.225\textwidth]{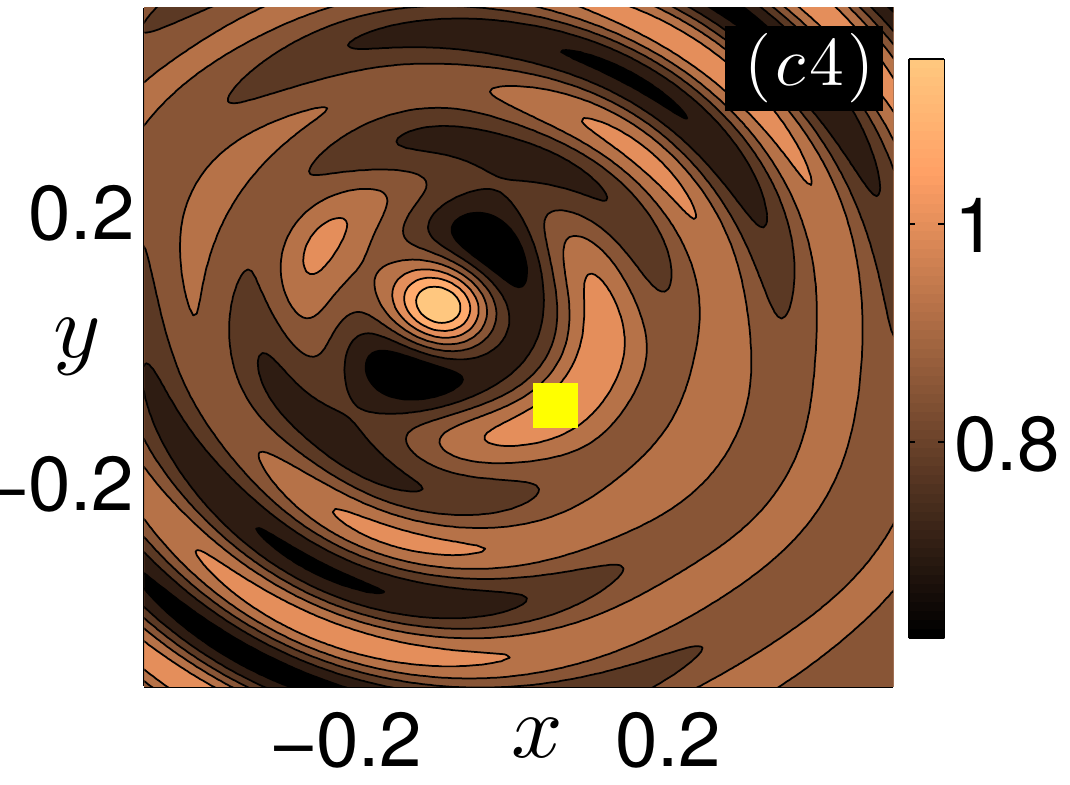} \\
\includegraphics[clip,width=0.225\textwidth]{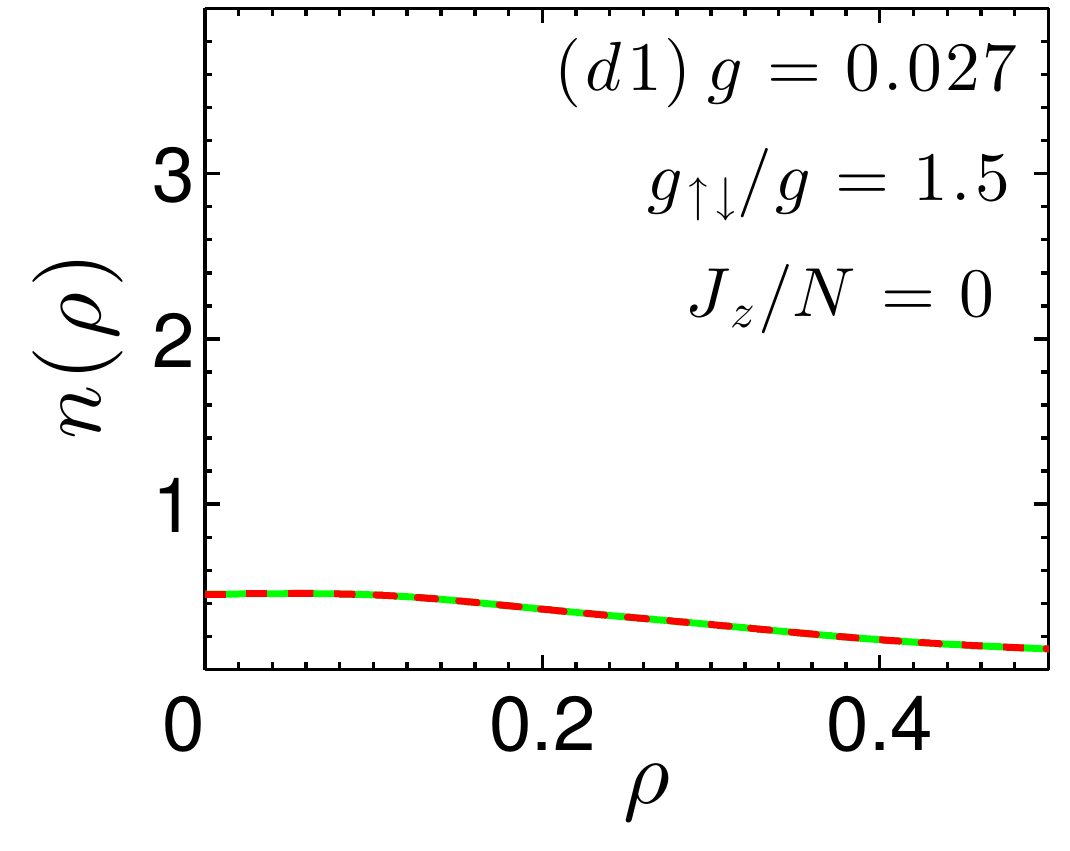} &
\includegraphics[clip,width=0.225\textwidth]{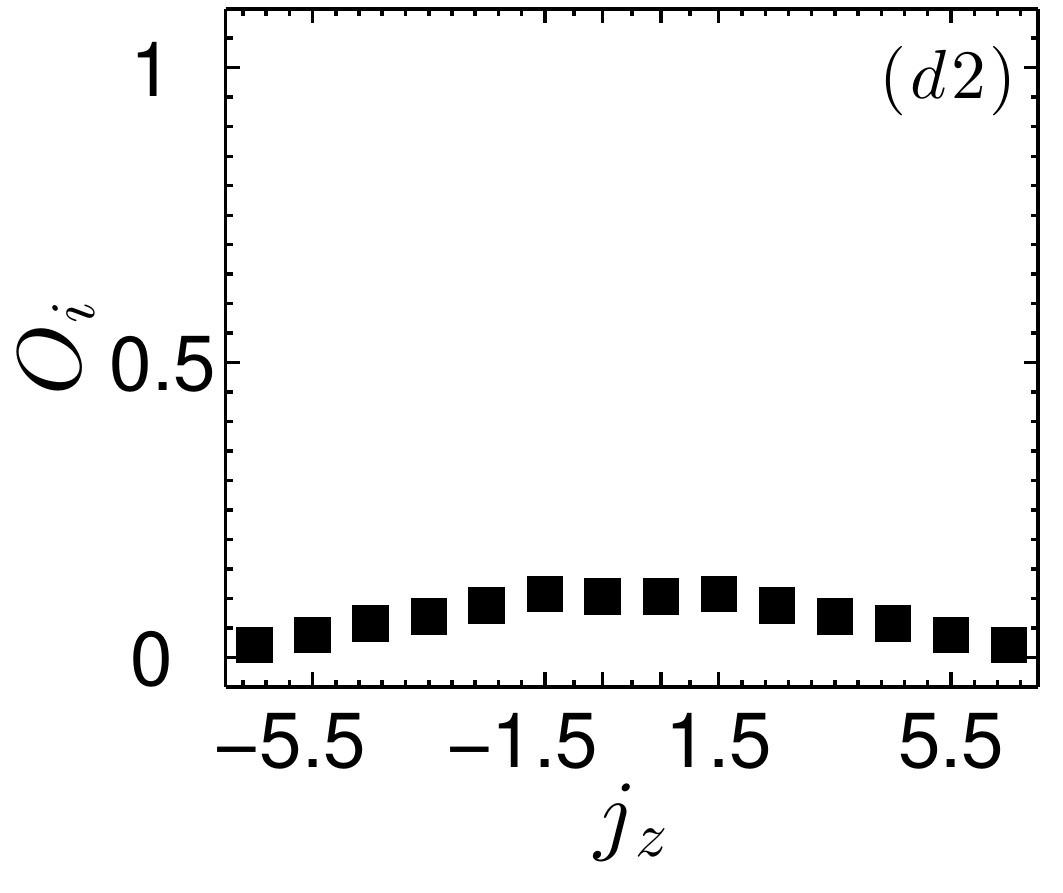} &
\includegraphics[clip,width=0.225\textwidth]{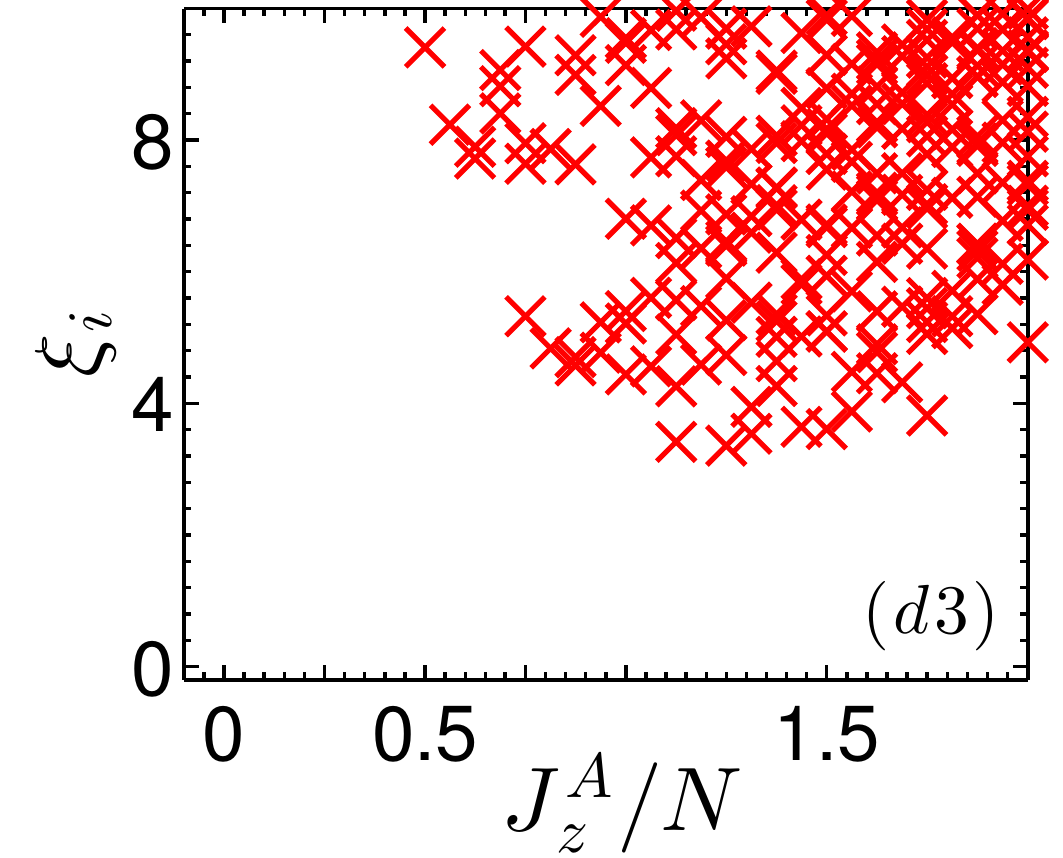} &
\includegraphics[clip,width=0.225\textwidth]{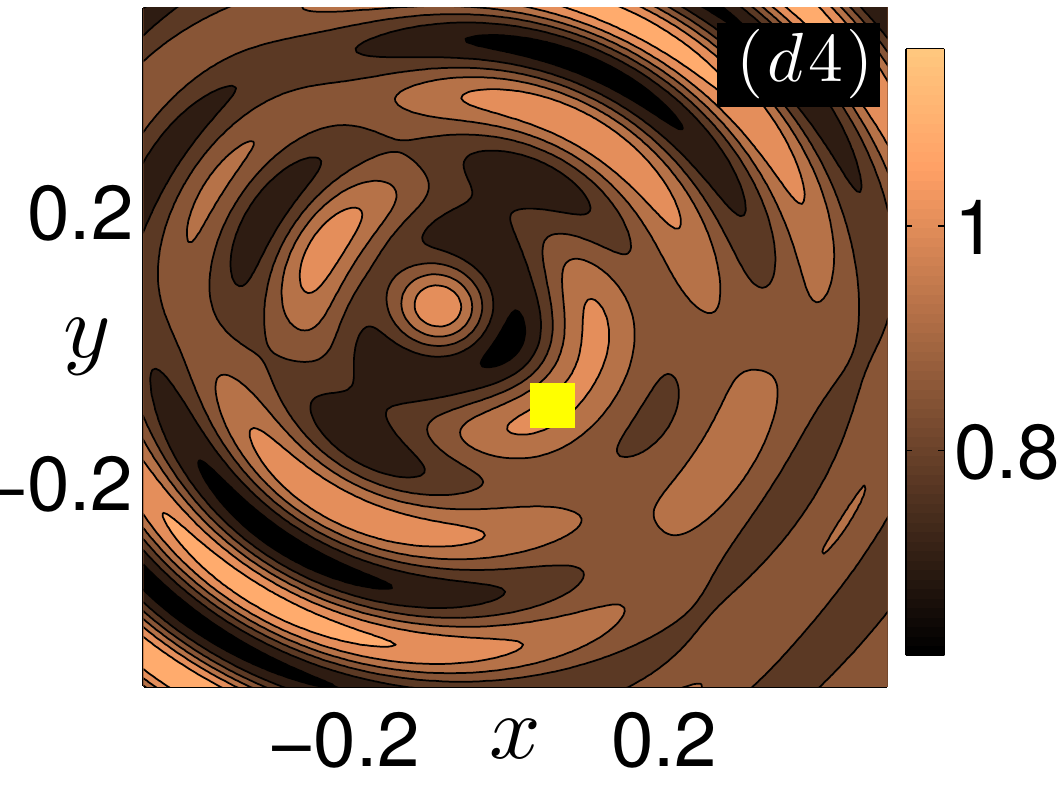} 
\end{tabular}
\caption{(color online). Plots in each row illustrate the ground state properties at a representative interaction strength in Fig.~\ref{figN8ee05}$(a)$ or  \ref{figN8ee15}$(a)$. In the first column (from left), we show density distributions of spin-up component $n_{\uparrow}(\rho)$ (solid green) and of spin-down component $n_{\downarrow}(\rho)$ (dashed red). In the second column, we show eigenvalues $O_i$ of single-particle density matrix as a function of angular momentum $j_z$ of the single-particle states $|\phi_i(\textbf{r})\rangle$. In the third column, we show corresponding $OES$ plots of entanglement pseudo-energies $\xi_i$ as a function of $J_z^A/N$, the average angular momentum of subsystem $A$. In the last column, we show contour plots $(a4)$, $(c4)$, and $(d4)$ that are normalized pair-correlation functions $\langle n_{\uparrow}(\textbf{r}_0) n_{\downarrow}(\textbf{r})\rangle$, with $\textbf{r}_0$ denoted by a (yellow) marker. Phase plot $(b4)$ is derived from reduced wavefunction $\psi_{c,\downarrow}({\bf r})$, which is computed by fixing 7 of the 8 particles at their most probable locations and their corresponding radii are indicated by (yellow) markers. Closed dashed (blue) contour is a guide to the eye, that allows us to count the number of phase slips.}
\label{figN8all}
\end{figure*}
\par
The detailed analysis presented above for the relatively simple, but rich, scenario of $N=2$ particles shall be useful when discussing results at larger particle numbers. Even with a small increase in particle number from $N=2$ to $N=4$ (not shown), we observe the non-occurrence of ground states in ${\cal P}4$ phase. As discussed in the introduction of Sec.~\ref{secresults}, this can be understood as a manifestation of the competition between energy contributions from ${\cal H}_0$ and ${\cal H}_\textrm{int}$. A higher particle number increases the probability distribution into single-particle states with smaller angular momenta, when compared to larger angular momenta eigenstates. We shall now proceed to consider the few-body system with $N=8$ particles, discuss the occurrence of various phases, and analyze the ground state properties at representative interaction strengths using various techniques outlined in Sec.~\ref{secanalysis}. 
\par
\textbf{Figs.~\ref{figN8ee05}($\boldsymbol{a}$), \ref{figN8ee15}($\boldsymbol{a}$):} We solve the interacting Hamiltonian $\cal H$ at various interaction strengths and identify corresponding ground state manifolds $J_z/N$ in Figs.~\ref{figN8ee05}$(a)$ and \ref{figN8ee15}$(a)$. As discussed with reference to Figs.~\ref{figN2ee05}($a$) and \ref{figN2ee15}($a$), it is evident that depending on $g$ and $g_{\uparrow\downarrow}$, the ground states belong to different $J_z/N$ manifolds, and in-turn to ${\cal PT}$ or ${\cal P}$ symmetry phases. In this relatively larger particle number scenario, we observe that the ground states fall into two distinct regimes: (a) at weak interaction strengths (\emph{mean-field-like regime}), we observe ground states with topological and symmetry properties that are consistent with mean-field theory computations \cite{ourPRL}; (b) at intermediate to strong interaction strengths (\emph{strongly correlated regime}), we report the emergence of strong correlations in ground states. The strongly correlated ground states are eigenstates of ${\cal PT}$ operator, and we additionally identify them with the label `$SC$'. In Fig.~\ref{figN8all}, we illustrate the ground state properties at representative interaction strengths in these two regimes. 
\par
\textbf{\emph{Mean-field-like regime:-} Figs.~\ref{figN8all}($\boldsymbol{a1}$) $\rightarrow$ \ref{figN8all}($\boldsymbol{a4}$), \ref{figN8all}($\boldsymbol{b1}$) $\rightarrow$ \ref{figN8all}($\boldsymbol{b4}$):} In the top row, we illustrate the ground state properties of the ${\cal PT}$ eigenstate in $J_z/N=0$ manifold at $g=0.001$ of Fig.~\ref{figN8ee05}$(a)$. It is evident that the properties in Figs.~\ref{figN8all}($a1$) $\rightarrow$ \ref{figN8all}($a4$) are qualitatively identical to their counterparts in Figs.~\ref{figN2all}($a1$) $\rightarrow$ \ref{figN2all}($a4$). In the second row, we discuss the ground state properties of the ${\cal P}$ eigenstate in $J_z/N=+1.5$ manifold at $g=0.013$ of Fig.~\ref{figN8ee05}$(a)$. The corresponding ground state is degenerate in $J_z/N= \pm1.5$ manifolds, while we restrict our discussion to $J_z/N= +1.5$ manifold. As expected for a ${\cal P}$ eigenstate, the density distributions $n_{\uparrow}(\rho)$ and $n_{\downarrow}(\rho)$ shown in Fig.~\ref{figN8all}$(b1)$ are distinct. It is evident from the single-particle density matrix eigenvalues in Fig.~\ref{figN8all}$(b2)$ that there is a peak in the occupation of eigenstate identified by $j_z = +1.5$. From the corresponding $OES$ plot in Fig.~\ref{figN8all}$(b3)$, we observe that the ground state is predominantly occupied by $\xi_i$ at $J_z^A/N=1.5$. To illustrate the internal structure of this ${\cal P}$ eigenstate, we show the phase plot derived from the reduced wavefunction $\psi_{c,\downarrow}({\bf r})$ in Fig.~\ref{figN8all}$(b4)$. It is evident from the representation in Eqn.~(\ref{eqnspstates}) that the orbital angular momentum of spin-up component in the ground state is +1 and that of spin-down component is +2. Correspondingly, the phase plot of the down-spin component shown in Fig.~\ref{figN8all}$(b4)$ exhibits a vorticity of 2, and hence we identify this ${\cal P}$ eigenstate as ${\cal P}2$. 
\par
\textbf{\emph{Strongly correlated regime:-} Figs.~\ref{figN8all}($\boldsymbol{c1}$) $\rightarrow$ \ref{figN8all}($\boldsymbol{c4}$), \ref{figN8all}($\boldsymbol{d1}$) $\rightarrow$ \ref{figN8all}($\boldsymbol{d4}$):} In the third and fourth rows, we illustrate the ground state properties of the ${\cal PT}$ eigenstates in the strongly correlated regime at $g=0.021$ of Fig.~\ref{figN8ee05}$(a)$ and $g=0.027$ of Fig.~\ref{figN8ee15}$(a)$ respectively. At intermediate to strong interaction strengths, as shown in Figs.~\ref{figN8ee05}$(a)$ and Fig.~\ref{figN8ee15}$(a)$, all the ground states in this regime are eigenstates of ${\cal PT}$ operator in $J_z/N=0$ manifold. As expected, the density distributions $n_{\uparrow}(\rho)$ and $n_{\downarrow}(\rho)$ overlap in Figs.~\ref{figN8all}$(c1)$ and \ref{figN8all}$(d1)$. We observe that the density distributions become increasingly flat with increasing magnitude of interaction strengths, $g$ and $g_{\uparrow\downarrow}$. The interaction-induced correlations present in the ground states are revealed by the eigenvalues of single-particle density matrix and $OES$ plots. From the plots in Figs.~\ref{figN8all}$(c2)$ and \ref{figN8all}$(d2)$, it is evident that the particles are nearly uniformly distributed across many single-particle eigenstates, with an equal distribution among time-reversal partner states. This distribution is qualitatively in the opposite limit to the corresponding plots in the mean-field-like regime illustrated in Figs.~\ref{figN8all}$(a2)$ and \ref{figN8all}$(b2)$. This feature is further substantiated in the $OES$ plots of Figs.~\ref{figN8all}$(c3)$ and \ref{figN8all}$(d3)$, where a large number of entanglement pseudo-energies $\xi_i$ are degenerate or \emph{nearly} degenerate. As discussed in Sec.~\ref{secem}, the presence of a large degeneracy in entanglement pseudo-energies is a clear manifestation of the strongly correlated nature of the ground states. We further observe that with increasing interaction strengths, the minima of the entanglement pseudo-energies $\xi_i$ shifts to larger $J_z^A/N$ values. To illustrate the internal structure and the correlations between up-spin and down-spin components of these ${\cal PT}$ eigenstates, we show the pair-correlation functions in Figs.~\ref{figN8all}$(c4)$ and \ref{figN8all}$(d4)$. 
\par
With our understanding of $OES$ plots in Figs.~\ref{figN8all}, we may now explain various features observed in $EE$ plots that help us understand the correlation properties of the ground states in the mean-field-like and strongly correlated regimes. As discussed with reference to Figs.~\ref{figN2ee05}($b$) and \ref{figN2ee15}($b$), we observe qualitatively similar features in $N=8$ particle case as well. The presence of distinctly different slopes in Figs.~\ref{figN8ee05}($b$) and \ref{figN8ee15}($b$) suggests the presence of distinct correlation properties in different ground states within various phases. Within each phase, $EE$ increases monotonously with increasing $g$ due to the presence of increased correlations in the ground state. For example, to illustrate this feature within the ${\cal PT} (SC)$ phase, we may compare $OES$ plots in Figs.~\ref{figN8all}$(c3)$ and \ref{figN8all}$(d3)$ and observe an increased homogeneity in Fock states. As a side note, we observe a small region of ${\cal P}$-symmetric states before the transition to strongly correlated regime. These states do not possess distinct topological or correlation properties. Without loss of generality, we assert that these ground states merely occupy a \emph{crossover} region prior to the transition to strongly correlated regime. 
\par
In summary, we emphasize that the ground states in the weakly interacting regime illustrated in the top two rows of Fig.~\ref{figN8all} are mean-field-like states. Their density distributions, pair-correlation functions and reduced wavefunctions may be readily related to the results from mean-field theory computations discussed in our earlier publication \cite{ourPRL}. Within the ED scheme, we even reproduce the reversal of phase symmetry between ${\cal P}$ and ${\cal PT}$ eigenstates that is observed with an increasing value of $g$, but with a fixed value of $g_{\uparrow\downarrow}/g$ in our earlier mean-field study \cite{ourPRL}. Such a correspondence between ED results and mean-field theory results is anticipated only when the ground state is predominantly occupied by one single-particle eigenstate (and/or its time-reversal partner), as revealed in Figs.~\ref{figN8all}$(a2)$ and \ref{figN8all}$(b2)$. As illustrated in the bottom two rows of Fig.~\ref{figN8all}, the presence of a large degeneracy in entanglement pseudo-energies and the distribution of particles across many single-particle eigenstates, are clear manifestations of the strongly correlated nature of the ground states. Furthermore, we observe from Figs.~\ref{figN8all}, that the transition from mean-field-like regime to a strongly correlated regime is attained with only small variations in the magnitudes of inter-particle interaction strengths. We emphasize here that the pivotal reason behind this feature is the presence of \emph{nearly flat} single-particle energy spectrum at large SO-coupling strengths.
\section{Conclusions}
\label{seccon}
We systematically study an interacting few-body system of two-component Bose gases with 2D isotropic Rashba SO-coupling in a 2D isotropic harmonic trap. We show that the model Hamiltonian is gauge-equivalent to particles subject to a $\cal T$-symmetry preserving pure non-abelian vector potential, whose magnitude proportionally determines the strength of Rashba SO-coupling. It is experimentally feasible to device a scheme in which tunable parameters, such as laser fields, can be used to control the magnitude of non-abelian vector potential, and hence simulate large SO-coupling strengths. In this limit of large SO-coupling strengths, we show that the single-particle energy spectrum is \emph{nearly flat}. In the recent past, several research groups have made proposals to engineer quantum systems in which interactions would play a dominant role and the ground states would in-turn be strongly correlated. For example, recent proposals suggest schemes that would engineer \emph{nearly flat} Chern bands to study strongly correlated fractional quantum Hall states in the lattice limit \cite{FQHflat}. Though we study few-body Bose gases in traps, we emphasize that the intention with which we have identified the existence of \emph{nearly flat} energy spectra at large SO-coupling strengths is not too dissimilar from the afore-mentioned line of thought. 
\par
In our model system with \emph{nearly flat} energy spectra, we observe that the presence of inter-particle interactions allows for the emergence of ground states with distinct topological, symmetry and correlation properties. We solve the interacting Hamiltonian in different particle number scenarios and analyze the ground state properties with the help of energy spectrum, single-particle density matrix, pair-correlation functions, reduced wavefunctions, and entanglement measures. At small particle numbers, we show the phase diagram in Figs.~\ref{figN2ee05} and \ref{figN2ee15}, with ground states being eigenstates of either ${\cal P}$ or ${\cal PT}$ operator. In Fig.~\ref{figN2all}, we illustrate the ground state properties at representative interaction strengths in various phases. We further assert that the bosons condense to an array of topological ${\cal P}$ eigenstates with $n+1/2$ quantum angular momentum vortex configuration, with $n = 0, 1, 2, 3,$. At large particle numbers, we illustrate the phase diagram in Figs.~\ref{figN8ee05} and \ref{figN8ee15}. We observe the presence of two distinct regimes: (a) at weak interaction strengths (mean-field-like regime), we obtain ground states with topological and symmetry properties that are also obtained via mean-field theory computations. We justify this correspondence and illustrate the ground state properties in detail in Fig.~\ref{figN8all}. (b) at intermediate to strong interaction strengths (strongly correlated regime), we report the emergence of strongly correlated ground states. The properties illustrated in Fig.~\ref{figN8all} demonstrate the correlated nature of the ground states. 
\par
It is interesting to inquire if the strongly correlated ground states that emerge in the \emph{nearly flat} energy spectra would eventually allow for the manifestation of bosonic analogues of topological insulators predicted to occur in traditional condensed matter systems. We emphasize that in our system of trapped bosons, quantum statistics makes it impossible to fill up the lowest generalized Landau level. This results in the absence of `sharp boundaries', which in-turn obviates the occurrence of states with topological order. However, this fundamental roadblock may be circumvented when we consider a system of SO-coupled bosons or fermions in specially engineered optical lattices \cite{SarmaTI}. 
\begin{acknowledgments}
RB thanks L. O. Baksmaty, C. Zhang, H. Lu and L. Dong for useful discussions. RB and HP acknowledge support by the NSF (PHY-1205973), the Welch Foundation (Grant No. C-1669) and the DARPA OLE program. HH was supported by the ARC Discovery Project DP0984522.
\end{acknowledgments}

\begin{thebibliography}{100}
%
\bibitem{review_trap_ole} I. Bloch, J. Dalibard, and W. Zwerger, Rev. Mod. Phys. \textbf{80}, 885 (2008);  M. Inguscio, W. Ketterle, and C. Salomon, \emph{Ultra-cold Fermi Gases} (IOS Press, Amsterdam, 2008); M. Lewenstein et al., Advances in Physics, \textbf{56}, 243 (2007).
%
\bibitem{SF} R. Onofrio et al., Phys. Rev. Lett. \textbf{85}, 2228 (2000); R. Desbuquois et al., Nature Physics \textbf{8}, 645 (2012); M. W. Zwierlein et al., Nature \textbf{435}, 1047 (2005).
%
\bibitem{Effimov} S. E. Pollack, D. Dries, and R. G. Hulet, Science \textbf{326}, 5960 (2009); J. R. Williams et al., Phys. Rev. Lett. \textbf{103}, 130404 (2009).
%
\bibitem{MI} M. Greiner et al., Nature \textbf{415}, 39 (2002); U. Schneider et al., \textbf{322}, 5907 (2008); R. J\"{o}rdens et al., Nature \textbf{455}, 204 (2008).
% 
\bibitem{AFM} P. M. Duarte et al., Phys. Rev. A \textbf{84}, 061406(R) (2011).
%
\bibitem{frustrated} J. Struck et al., Science \textbf{333}, 996 (2011).
%
\bibitem{IBSemf} Y.-J. Lin et al., Nature \textbf{462}, 628 (2009); Y.-J. Lin et al., Nature Physics \textbf{7}, 531 (2011).
%
\bibitem{gaugefields_review} J. Dalibard et al., Rev. Mod. Phys. \textbf{83}, 1523 (2011).
%
\bibitem{SpielmanNature2011} Y.-J. Lin et al., Nature \textbf{471}, 83 (2011).
%
\bibitem{gauge_fermions} P. Wang et al., Phys. Rev. Lett. \textbf{109}, 095301 (2012); Lawrence W. Cheuk et al., Phys. Rev. Lett. \textbf{109}, 095302 (2012).
%
\bibitem{theory_soc_TI} C. L. Kane and E. J. Mele, Phys. Rev. Lett. \textbf{95}, 226801 (2005); B. A. Bernevig and S.-C. Zhang, Phys. Rev. Lett. \textbf{96}, 106802 (2006). 
%
\bibitem{IQH} M. Burrello and A. Trombettoni, Phys. Rev. A \textbf{84}, 043625 (2011); B. Juli\'{a}-D\'{i}az et al., New J. Phys \textbf{14}, 055003 (2012).
%
\bibitem{PetrovPRL2000} D. S. Petrov, M. Holzmann, and G. V. Shlyapnikov,
Phys. Rev. Lett. \textbf{84}, 2551 (2000).
%
\bibitem{Dalibard2D} T. Yefsah, R. Desbuquois, L. Chomaz, K. J. Gunter, and J. Dalibard, Phys. Rev. Lett. \textbf{107}, 130401 (2011).
%
\bibitem{gaugeprop1} M. Burrello and A. Trombettoni, Phys. Rev. Lett. \textbf{105}, 125304 (2010).
%
\bibitem{LLwu} Y. Li, X. Zhou, and C. Wu, Phys. Rev. B \textbf{85}, 125122 (2012).
%
\bibitem{exp_dyn} Z. F. Xu and L. You, Phys. Rev. A \textbf{85}, 043605 (2012).
% 
\bibitem{HV12pra} B. Ramachandhran et al., Phys. Rev. A \textbf{85}, 023606 (2012).
%
\bibitem{EDpaper} J. M. Zhang and R. X. Dong, Eur. J. Phys. \textbf{31}, 591 (2010): techniques illustrated in this manuscript were particularly useful during implementation. 
%
\bibitem{Tsymm} L. O. Baksmaty, C. Yannouleas, and U. Landman, Phys. Rev. A \textbf{75}, 023620 (2007); B. Juli\'{a} - D\'{i}az, D. Dagnino, K. J. Gunter, T. Grass, N. Barberan, M. Lewenstein, and J. Dalibard, Phys. Rev. A \textbf{84}, 053605 (2011).
%
\bibitem{Lewenstein06} N. Barber\'{a}n, M. Lewenstein, K. Osterloh, and D. Dagnino, Phys. Rev. A \textbf{73}, 063623 (2006).
%
\bibitem{CWF} H. Saarikoski et al.,  Europhys. Lett. \textbf{91}, 30006 (2010). 
%
\bibitem{ES} Y. Shi, Phys. Rev. A \textbf{67}, 024301 (2003); Y. Shi, J. Phys. A: Math. Gen. \textbf{37}, 6807 (2004).
% 
\bibitem{OES} A. Sterdyniak, B. A. Bernevig, N. Regnault, and F. D. M. Haldane, New J. Phys. \textbf{13}, 105001 (2011).
%
\bibitem{Haldane} H. Li and F. D. M. Haldane, Phys. Rev. Lett. \textbf{101}, 010504 (2008).
%
\bibitem{HaqueEE} O. S. Zozulya, M. Haque, and N. Regnault, Phys. Rev. B \textbf{79}, 045409 (2009). 
%
\bibitem{WuCPL2011} C. Wu, I. Mondragon-Shem, and X.-F. Zhou, Chin.
Phys. Lett. \textbf{28}, 097102 (2011).
%
\bibitem{HQVS} M. M. Salomaa and G. E. Volovik, Phys. Rev. Lett.
\textbf{55}, 1184 (1985).
%
\bibitem{quadrops} N. Gemelke, E. Sarajlic, and S. Chu, arXiv: 1007.2677; H. Saarikoski, S. M. Reimann, A. Harju , and M. Manninen, Rev. Mod. Phys. \textbf{82}, 2785 (2010).
%
\bibitem{ZhaiPRL2010} C. Wang, C. Gao, C.-M. Jian, and H. Zhai, Phys.
Rev. Lett. \textbf{105}, 160403 (2010).
%
\bibitem{ourPRL} H. Hu, B. Ramachandhran, H. Pu, and X.-J. Liu,  Phys. Rev. Lett. \textbf{108}, 010402 (2012).
%
\bibitem{comments1} In Fig.~\ref{figN2ee15}$(b)$, ${\cal P}2 \rightarrow {\cal P}3$ transition neither exhibits a change in slope nor a noticeable drop in $EE$ value. From an analysis of $OES$ plots across this transition (not shown), we observe that the maximally contributing entanglement pseudo-energies $\xi_i$ at $J_z^A=1.5$ (${\cal P}2$) and $J_z^A=2.5$ (${\cal P}3$) are nearly degenerate, and hence we observe this anomaly. In a broader sense, we conclude that the ground states with ${\cal P}2$ symmetry in Fig.~\ref{figN2ee15}$(b)$ may merely occupy a small \emph{crossover} region between ${\cal PT}$ and ${\cal P}3$ phases. 
%
\bibitem{FQHflat} R. Roy and S. L. Sondhi, Physics \textbf{4}, 46 (2011); T. Neupert, L. Santos, C. Chamon, and C. Mudry, Phys. Rev. Lett. \textbf{106}, 236804 (2011);  E. Tang, J. -W. Mei, and X. -G. Wen, \emph{ibid}. \textbf{106}, 236802 (2011); K. Sun, Z. Gu, H. Katsura, and S. DasSarma, \emph{ibid}. \textbf{106}, 236803 (2011).
%
\bibitem{SarmaTI} T. D. Stanescu, V. Galitski, J. Y. Vaishnav, C. W. Clark, and S. DasSarma, Phys. Rev. A \textbf{79}, 053639 (2009): In this reference, the authors propose to realize topological phases emerging from single-particle Hamiltonian in optical lattices. 
%
%
\end{thebibliography}
\end{document}